\RequirePackage{silence}
\documentclass[fleqn,usenatbib,useAMS]{mnras} 
\usepackage{graphicx}
\usepackage{amsmath}

\usepackage{amsfonts}
\usepackage{float}
\usepackage{bm}
\usepackage[perpage,symbol*]{footmisc}
\usepackage{ae,aecompl}
\usepackage{array}
\usepackage{soul}
\usepackage{mathtools}
\usepackage{multirow}
\usepackage{rotating}

\defcitealias{Pointecouteau_2005}{PS05}

\makeatletter

\DeclareRobustCommand{\Rmnum}[1]{\expandafter\@slowromancap\romannumeral #1@}
\makeatother

\title[Hydrodynamical structure formation in MOND]{Hydrodynamical structure formation in Milgromian cosmology} 

\author[N. Wittenburg et al.]{\parbox[t]{\textwidth} {Nils Wittenburg$^{1}$\thanks{Email:
\href{mailto:nwittenburg@astro.uni-bonn.de}{nwittenburg@astro.uni-bonn.de} (Nils Wittenburg), \newline $~~~~~~~~~~~~~~~\,$ \href{mailto:pkroupa@uni-bonn.de}{pkroupa@uni-bonn.de} (Pavel Kroupa), \newline $~~~~~~~~~~~~~~~\,$ \href{mailto:indranilbanik1992@gmail.com}{indranilbanik1992@gmail.com} (Indranil Banik)}, Pavel Kroupa$^{1,2}$, Indranil Banik$^{3,1}$, Graeme Candlish$^4$\\
and Nick Samaras$^{2}$} \vspace{10pt}\\
$^{1}$Helmholtz-Institut f\"ur Strahlen und Kernphysik (HISKP), University of Bonn, Nussallee 14$-$16, D-53115 Bonn, Germany\\
$^{2}$Astronomical Institute, Faculty of Mathematics and Physics, Charles University, V Hole\v{s}ovi\v{c}k\'ach 2, CZ-180 00 Praha 8, Czech Republic\\
$^{3}$Scottish Universities Physics Alliance, University of Saint Andrews, North Haugh, Saint Andrews, Fife, KY16 9SS, UK\\
$^4$Instituto de F\'{i}sica y Astronom\'{i}a, Universidad de Valpara\'{i}so, Gran Breta\~{n}a 1111, 2360102 Valpara\'{i}so, Chile}

\pubyear{2023}
\begin{document}

\label{firstpage}
\pagerange{\pageref{firstpage}--\pageref{lastpage}}

\maketitle

\begin{abstract}  
We present the first hydrodynamical cosmological simulations in the $\nu$HDM framework based on Milgromian dynamics (MOND) with light (11~eV) sterile neutrinos. $\nu$HDM can explain the expansion history, CMB anisotropies, and galaxy cluster dynamics similarly to standard cosmology while preserving MOND's successes on galaxy scales, making this the most conservative Milgromian framework. We generate initial conditions including sterile neutrinos using \textsc{camb} and \textsc{music} and modify the publicly available code \textsc{phantom of ramses} to run $\nu$HDM models. The simulations start at redshift $z_e=199$, when the gravitational fields are stronger than $a_{_0}$ provided this does not vary. We analyse the growth of structure and investigate the impact of resolution and box size, which is at most 600 comoving Mpc. Large density contrasts arise at late times, which may explain the KBC void and Hubble tension. We quantify the mass function of formed structures at different redshifts. We show that the sterile neutrino mass fraction in these structures is similar to the cosmic fraction at high masses (consistent with MOND dynamical analyses) but approaches zero at lower masses, as expected for galaxies. We also identify structures with a low peculiar velocity comparable to the Local Group, but these are rare. The onset of group/cluster scale structure formation at $z_e\approx4$ appears to be in tension with observations of high redshift galaxies, which we discuss in comparison to prior analytical work in a MONDian framework. The formation of a cosmic web of filaments and voids demonstrates that this is not unique to standard Einstein/Newton-based cosmology.
\end{abstract}


\section{Introduction}
\label{Sec:Introduction}

Galaxies, groups of galaxies, and galaxy clusters form and evolve under cosmological boundary conditions which define their seed mass, interactions with each other, and accretion of gas from their surroundings. Most available self-consistent simulations of these processes have been made in the standard model of cosmology (SMoC), which assumes the universal validity of Einstein's general theory of relativity. To match observations of the expanding Universe, the cosmic microwave background (CMB) anisotropies, and the properties of nearby galaxies, the SMoC needs to be augmented by the auxiliary hypotheses that dark matter particles and dark energy dominate the matter and energy content of the Universe. Although the latter can be accounted for by a cosmological constant $\Lambda$ which is already a part of General Relativity, modified gravity theories \citep[e.g.,][]{Clifton_2012, Baker_2021} are usually considered in simulations only as an alternative to $\Lambda$ \citep[e.g.,][]{Li_2012}. Such models rarely question the hypothesis that most of the matter in the Universe and in galaxies is made of exotic dark matter particles which interact gravitationally but not electromagnetically and that are not accounted for in the standard model of particle physics. The exotic dark matter is generally assumed to consist of cold dark matter (CDM) particles that are non-relativistic at decoupling and interact only gravitationally with baryons, which combined with the cosmological constant and various other assumptions leads to the Lambda cold dark matter ($\Lambda$CDM) paradigm \citep*{Efstathiou_1990, Ostriker_1995}.

The $\Lambda$CDM simulations that have been and are being performed include, among many others, Illustris \citep{Vogelsberger_2014, Nelson_2015, Nelson_2019, Pillepich_2018}, EAGLE \citep{Schaye_2015, McAlpine_2016}, FIRE \citep{Hopkins_2014, Hopkins_2018}, HORIZON-AGN \citep{Dubois_2014}, and NEWHORIZON \citep{Dubois_2021}. These are based on different codes to treat hydrodynamics and different sub-grid algorithms to account for star formation and gas heating. The CDM part of the paradigm is sometimes altered by postulating different properties for the dark matter, assuming it to be, e.g., fuzzy \citep[FDM;][]{Hui_2017, May_2021}, warm \citep[WDM; e.g.,][]{Lovell_2020}, or self-interacting \citep[SIDM; e.g.,][]{Rocha_2013}. While this alters the small-scale behaviour of the dark matter, constraints are often so tight that these alternative universes regularly end up looking like CDM models \citep{Rogers_2021}. For example, FDM would need to behave like CDM in order to explain the large Newtonian dynamical mass-to-light ratios of the smallest ultra-faint galaxies, while on large scales the behaviour is similar to CDM such that the large-scale structure would be similar to the $\Lambda$CDM model.

Simulations like those mentioned above have allowed testing the SMoC against observations to unprecedented depth, thereby uncovering problems which pertain to small-scale issues \citep[e.g.,][and references therein]{Kroupa_2010}, including the planes of satellites problem \citep{Kroupa_2005}, which seems particularly robust to how the baryonic physics is treated \citep{Pawlowski_2018, Pawlowski_2021_Nature_Astronomy, Pawlowski_2021}. There are also serious problems on Gpc scales \citep{Haslbauer_2020, Valentino_2020_flat, Migkas_2021, Mohayaee_2021, Secrest_2022, Bargiacchi_2023, Lenart_2023}. The El Gordo interacting galaxy cluster at redshift $z_e = 0.87$ causes by itself a $>6\sigma$ falsification of the SMoC \citep*{Asencio_2021}, while the Hubble tension (\citealp{Haslbauer_2020}; see also the review by \citealp{Valentino_2021}) is similarly serious and similarly immune to resolution through improved treatment of baryonic physics on galaxy scales. Furthermore, a direct test for the existence of CDM particles on galaxy scales using Chandrasekhar dynamical friction suggests these particles do not exist at $13\sigma$ confidence based on galaxy bars not slowing down as expected \citep{Roshan_2021_bar_speed}, while the past orbital dynamics of the M81 group also precludes solutions with dark matter haloes \citep*{Oehm_2017}. The observed perturbations to dwarf galaxies in the Fornax galaxy cluster and the lack of low surface brightness dwarfs towards its centre also lead to the same conclusion, this time using tidal stability arguments \citep{Asencio_2022}. For a comprehensive recent review of problems faced by $\Lambda$CDM, we refer the reader to \citet{Kroupa_2015} and \citet{Perivolaropoulos_2022}.

Given these tensions between theory and observations, it would be of great value to have available simulations of structure formation in a fundamentally different theoretical framework. Such simulations would allow us to better appreciate the above-mentioned tensions of the SMoC. Depending on the performance of such a model, we would obtain information as to whether the particular approach is promising or should be discarded altogether. Currently, the only other theoretical approach to cosmological structure formation that can plausibly address these issues and be coded in the form of particle equations of motion is based on Milgromian dynamics \citep[MOND;][]{Milgrom_1983, Bekenstein_1984, Famaey_McGaugh_2012, Merritt_2020, Banik_Zhao_2022}. The basic idea of MOND is to not restrict oneself to the Solar System-scale constraints on the behaviour of gravity that allowed Newton and Einstein to formulate their theories, but to also consider more recent results from the rotation curves of galaxies, which reveal a much richer dynamics \citep[][and references therein]{Faber_1979}. To explain these observations using only the directly detected baryonic mass in stars and gas, \citet{Milgrom_1983} conjectured that the gravitational acceleration $g$ in an isolated spherically symmetric system is asymptotically related to the Newtonian gravitational acceleration $g_{_N}$ of the baryons alone according to
\begin{eqnarray}
	g \to \begin{cases}
	g_{_N} \, , & \textrm{if} ~g_{_N} \gg a_{_0} \, , \\
	\sqrt{a_{_0}g_{_N}} \, , & \textrm{if} ~g_{_N} \ll a_{_0} \, .
	\end{cases}
	\label{g_cases}
\end{eqnarray}
The key new ingredient is a fundamental acceleration scale $a_{_0} = 1.2 \times 10^{-10}$~m/s\textsuperscript{2} $\approx 3.9$~pc/Myr\textsuperscript{2} \citep[this value is empirical but has remained stable for many decades; see][]{Begeman_1991, Gentile_2011, McGaugh_Lelli_2016}. This behaviour follows from a Lagrangian, ensuring the usual symmetries and conservation laws with respect to the linear and angular momentum and the energy \citep{Bekenstein_1984, Milgrom_2010}. The least action principle then leads to a generalised Poisson equation that is non-linear in the mass distribution.

Though we use a non-relativistic formulation in this work assuming a standard background expansion history, relativistic formulations exist which are consistent with the CMB anisotropies, strong gravitational lensing, and the fact that gravitational and electromagnetic waves travel at the same speed to high precision \citep{Skordis_2019, Skordis_2021, Skordis_2022}. MOND may follow from properties of the quantum vacuum in the presence of a cosmological constant \citep{Milgrom_1999}.

MOND has been uncannily successful, with all \emph{a priori} predictions on galactic scales made in \citet{Milgrom_1983} having been verified by subsequent observations (for reviews, see \citealt{Famaey_McGaugh_2012} and \citealt{Banik_Zhao_2022}). The first cosmological structure formation simulations that show the formation of a cosmic web were done by \citet{Llinares_2008}, where the authors assumed a purely baryonic universe with $\Omega_m=\Omega_b=0.04$ and treated the matter as pressureless dust, neglecting hydrodynamics. The problems of MOND on galaxy cluster scales can be addressed without needing the specific features of the scalar and vector field of the \citet{Skordis_2021} model by postulating sterile neutrinos with a rest energy of 11~eV \citep{Angus_2009, Angus_2011, Angus_2013}. Originally the idea of adding a collisionless component to account for the missing mass in MOND on cluster scales stems from \citet{Sanders_2003}, who argued at the time that ordinary neutrinos could be a possible candidate \citep[though this is no longer possible; see][]{Katrin_2019, Katrin_2022}. These 11~eV sterile neutrinos naturally account for the CMB similarly to $\Lambda$CDM because they have the same relic mass density as the CDM in the $\Lambda$CDM paradigm, leading to a standard expansion history and very similar physics in the early universe \citep[for a more detailed description, see section 3.1 of][]{Haslbauer_2020}. We stress here that the collisionless sterile neutrino component is not related to the exotic CDM particles in the SMoC because the postulated light sterile neutrinos would not cluster on galaxy scales \citep{Angus_2010_minimum_neutrino_mass}, though the cosmic mean density of the two is about the same.

This neutrino hot dark matter ($\nu$HDM) model\footnote{The postulated collisionless sterile neutrinos are not related to the exotic dark matter particles in the SMoC, being instead motivated from neutrino flavour oscillations observed in terrestrial experiments \citep{Merle_2017}.} was implemented numerically by \citet{Angus_2011} and later by \citet{Katz_2013}, while \citet{Haslbauer_2020} used semi-analytic methods to show that it mitigates the Hubble tension by means of galaxies falling towards the void edge during the formation of a large local supervoid, which is indeed observed as the KBC void \citep*{Keenan_2013}. Turning to overdensities, \citet{Asencio_2021} used the $N$-body $\nu$HDM simulations of \citet{Katz_2013} to show that this framework also accounts for the El Gordo galaxy cluster collision, thereby producing roughly the right frequency of very massive interacting galaxy clusters at high $z_e$, a test which $\Lambda$CDM fails with $>6\sigma$ confidence. While the \citet{Llinares_2008} and the \citet{Katz_2013} simulations considered only large scales and thus treated the baryons as dissipationless dust, currently no equivalent modelling exists which takes into account the hydrodynamics in order to resolve scales down to individual galaxies.

We simulate 300 and 600 comoving Mpc (cMpc)-scale cosmological volumes with gas and sterile neutrinos, starting at $z_e = 199$ and evolved until the present time. This volume can reach galaxy group/cluster scales in moderately powerful in-house computer servers, thereby allowing a comparison with observed galaxy clusters. We introduce the publicly available method and codes\footnote{\url{https://bitbucket.org/SrikanthTN/bonnpor/src/master/Cosmo_patch_and_setup_and_halofinder/}} used here and analyse the mass function of the resulting galaxy cluster population at different epochs. An important aspect of our simulated massive galaxy clusters is that they are dominated by sterile neutrinos, thus alleviating concerns about the application of MOND to the Bullet Cluster \citep{Angus_2007}. Subsequent publications will report more extensive analyses of the properties of galaxy groups and clusters down to individual galaxies, a scale which we are already able to reach in this work.

Section \ref{Sec:theory} introduces Milgromian gravitation and describes the $\nu$HDM cosmological framework. Since star formation is not taken into account in this first exploratory modelling, the method to find bound structures is described in Section~\ref{Sec:halofinder}. The models computed here are detailed in Section~\ref{Sec:models} and their results are discussed in Section~\ref{Sec:results}. Section~\ref{Sec:conclusion} concludes this contribution. Movies of the simulations are publicly available.\footnote{\url{https://www.youtube.com/playlist?list=PL2mtDSIH4RQhtoIxlOBvDzrzhT7bAFxqq}} We emphasize that these are the first ever hydrodynamic simulations performed in a MOND cosmological model. The here-studied $\nu$HDM model is conservative in the sense that it retains a very early inflationary phase, assumes the FLRW metric to be valid, and assumes dark energy, allowing it to reproduce the CMB with its high third peak and yield the same expansion history as the SMoC.

\section{Theoretical background and numerical methods}
\label{Sec:theory}

\subsection{MOND}

MOND \citep{Milgrom_1983} can be formulated as a non-relativistic theory of gravity with a fully developed Lagrangian formalism \citep{Bekenstein_1984, Famaey_McGaugh_2012}. MOND was originally developed by \citet{Milgrom_1983} to explain the missing gravity problem on galaxy scales \citep[evidenced in their flat outer rotation curves; see][and references therein]{Bosma_1978, Rubin_1978, Faber_1979} without invoking the existence of exotic dark matter particles. The underlying reason for the proposed departure from standard physics may be related to the quantum physics of the vacuum \citep*{Milgrom_1999, Pazy_2013, Smolin_2017, Senay_2021}. According to MOND, the gravitational accelerations depart from the Newtonian behaviour in the MOND regime of very weak gravitational fields ($g \ll a_{_0}$), but standard dynamics is recovered in the Newtonian or strong-field regime ($g \gg a_{_0}$). Equation~\ref{g_cases} shows the asymptotic regimes, the transition between which occurs when $g$ is comparable to Milgrom's constant $a_{_0}$. To transition between these regimes and handle more complicated geometries while also basing the theory on a Lagrangian to ensure the usual symmetries and conservation laws, \citet{Bekenstein_1984} developed the following MOND Poisson equation:
\begin{eqnarray}
    \nabla \cdot \left[ \mu\left( \frac{\lvert \nabla \Phi \rvert}{a_{_0}} \right) \nabla\Phi \left( \bm{r} \right) \right] = 4\mathrm{\pi} G\rho\left( \bm{r} \right) \equiv \nabla \cdot \nabla \Phi_N \left( \bm{r} \right) \, ,
    \label{eq:PoissonMOND}
\end{eqnarray}
where $\Phi$ is the potential, $N$ subscripts denote Newtonian quantities, $\bm{r}$ is the position, and $\mu$ is the interpolating function with dimensionless argument $y_\mathrm{int} \equiv \lvert \nabla \Phi \rvert/a_{_0}$, the gravitational field in units of $a_{_0}$. For consistency with the empirical Equation~\ref{g_cases},
\begin{eqnarray}
	\mu \left( y_\mathrm{int} \right) \to \begin{cases}
	y_\mathrm{int} \, , & \textrm{if} ~g \ll a_{_0}  \left( y_\mathrm{int} \ll 1 \right) \, , \\
	1 \, , & \textrm{if} ~g \gg a_{_0} \left( y_\mathrm{int} \gg 1 \right) \, .
	\end{cases}
	\label{mu_cases}
\end{eqnarray}
In spherical symmetry, $\mu \bm{g} = \bm{g}_{_N}$. Equation~\ref{eq:PoissonMOND} was commonly used in past simulations \citep[e.g., see the publicly available numerical implementation in \textsc{raymond};][]{Candlish_2015}. Nowadays, a computationally less intensive formulation is typically implemented which has the same asymptotic behaviour (see Section~\ref{Sec:simulationcode}).

One of the most valuable attributes of MOND is its highly predictive nature on galaxy scales \citep{Merritt_2020}. In particular, MOND predicted a tight baryonic Tully-Fisher relation with slope 4 and anticipated that the relation would be tightest between the flat part of the rotation curve and the total baryonic mass \citep{Lelli_2019}. Observations show that considering other velocity scales like the maximum of the rotation curve leads to a weaker correlation, while considering only the stellar mass rather than the total baryonic mass also leads to a breakdown of the relation at the low-mass end where the contribution of gas is important \citep{McGaugh_2012}. Beyond just the flat outer part of the rotation curve, MOND predicted a tight radial acceleration relation (RAR)\footnote{Called the mass discrepancy-acceleration relation (MDAR) in earlier works.} between the observed $g$ and $g_{_N}$, which is very apparent in the observational data \citep{Sanders_1990, McGaugh_2004, McGaugh_2005, McGaugh_2012, Wu_2015, Lelli_2017}, including lenticular galaxies \citep{Shelest_2020}. For reviews of this strong empirical correlation, we refer the reader to \citet{Famaey_McGaugh_2012} and \citet{McGaugh_2020}. While the focus in the past has been on rotating disc galaxies, MOND has also been successfully applied to predict the velocity dispersions of isolated dwarf galaxies in the Local Group \citep{McGaugh_2021}. Indeed, it has been argued in an extensive recent review that the RAR is ``not unique to rotationally supported systems or to rotational motion, nor is it confined to thin disc galaxies'' \citep[section~3 of][]{Banik_Zhao_2022}. The RAR also extends to weak gravitational lensing, which demonstrates its validity down to $g_{_N} \approx 10^{-5} \, a_{_0}$ \citep{Brouwer_2021}.

\subsection{The \texorpdfstring{$\nu$}{v}HDM model}
\label{Sec:numodel}

Despite the impressive successes of MOND on galaxy scales, it cannot easily account for observations on larger scales like the dynamics of galaxy clusters \citep{Sanders_1999, Sanders_2003} and the CMB anisotropies \citep{McGaugh_2004, Planck_2020} if we restrict ourselves to a purely baryonic universe. One recent proposal to address this issue has been the new aether scalar tensor (AEST) relativistic theory of \citet{Skordis_2021} in which the action has a function that plays the role of the MOND interpolating function, but its value depends on both the spatial gradient squared of the scalar field and its temporal derivative. The time-dependent term behaves like gravitating dust, allowing AEST to reproduce the angular power spectrum of the CMB and the $\Lambda$CDM result for the matter power spectrum on large scales. One can speculate that it might also give rise to additional gravitational acceleration inside galaxy clusters, explaining the residual missing mass of MOND \citep{Ettori_2019}. However, choosing AEST model parameters which might achieve this seems to cause problems matching the weak lensing RAR \citep*{Mistele_2023}. In any case, it is clear that significant fine-tuning would be required to consistently explain the dynamics of both galaxies and galaxy clusters with just their luminous mass.

A more conservative approach (Section~\ref{Sec:Introduction}) is to consider that it is perfectly possible to have both MONDian gravity and a type of collisionless matter that does not affect the MOND fits to galaxy rotation curves. \citet{Angus_2009} developed such a model by postulating the existence of sterile neutrinos with a mass of $m_\mathrm{\nu s} = 11$~eV/$c^2$, where $c$ is the speed of light. Thermal sterile neutrinos with this mass would have the same relic mass density as the CDM particles in $\Lambda$CDM: if half of all quantum states are occupied (similarly to the active neutrinos), no fine-tuning of their abundance would be necessary \citep[see][and references therein]{Diaferio_2012}. Combined with the high accelerations around the epoch of recombination ($g \approx 20 \, a_{_0}$) and an almost standard expansion history, this would lead to the early universe behaving similarly to $\Lambda$CDM \citep[for a more detailed discussion, see section~3.1 of][]{Haslbauer_2020}. The primordial abundances of light elements would then work much the same in both paradigms, with only small differences. The dynamics of galaxy clusters would also be accounted for in $\nu$HDM \citep*{Angus_2010}, including the offset between the weak lensing and X-ray peaks in the Bullet Cluster \citep{Angus_2007}.

From a particle physics perspective, the fact that ordinary neutrinos interchange their masses as they propagate suggests that there is an additional (sterile) neutrino in order to conserve kinetic energy and momentum. Sterile neutrinos are thus not equivalent to cold or fuzzy dark matter particles, but were in the first instance hypothesized to exist based on the observed properties of standard particles \citep{Merle_2017}. At a rest energy $>10$~eV, the free streaming length of sterile neutrinos would be short enough to not discernibly affect the CMB \citep[section~6.4.3 of][]{Planck_2016}. The Tremaine-Gunn limit to their phase space density \citep{Tremaine_Gunn_1979} would also be low enough that the dynamics of galaxies would not be affected by the sterile neutrinos, so galaxies would still be purely baryonic and MONDian \citep{Angus_2010_minimum_neutrino_mass}. This also implies that although $\nu$HDM contains a dominant collisionless matter component like in $\Lambda$CDM, it does not promote the formation of galaxies because it is unable to efficiently cluster on very small scales. We adopt this $\nu$HDM model and implement it into the simulation code \textsc{phantom of ramses} \citep[\textsc{por};][]{Lughausen_2015, Nagesh_2021}.

\subsection{Simulation code}
\label{Sec:simulationcode}

\textsc{por} is a customized version of the hydrodynamical $N$-body adaptive mesh refinement (AMR) code \textsc{ramses} \citep{Teyssier_2002} which implements the quasi-linear formulation of MOND \citep[QUMOND;][]{Milgrom_2010}. The novel approach is to interpret the Milgromian potential $\Phi \left( \bm{r} \right)$ as arising from the application of Newtonian gravity to some effective density distribution
\begin{eqnarray}
    \rho_{\textrm{eff}} ~\equiv~ \rho + \rho_p \, ,
\end{eqnarray}
where $\rho_p$ is the `phantom dark matter' (PDM) density defined such that $\nabla^2 \Phi \equiv 4\mathrm{\pi}G\rho_{\textrm{eff}}$. The PDM is merely a mathematical construction that does not correspond to real particles and therefore does not lead to Chandrasekhar dynamical friction. The advantage of thinking in this way is that whereas Equation~\ref{eq:PoissonMOND} involves a non-linear Poisson equation that is quite difficult to solve numerically, we may instead solve the following system of equations:
\begin{eqnarray}
    \label{eq:QUMOND}
    \!\!\!\!\!\!\nabla^2 \Phi \left( \bm{r} \right) &=& 4\mathrm{\pi} G\left[ \rho \left( \bm{r} \right) + \rho_p \left( \bm{r} \right) \right] \\
    &=& 4\mathrm{\pi} G\rho \left( \bm{r} \right) + \nabla \cdot \left[\widetilde{\nu} \left( \frac{\lvert \nabla \Phi_N \rvert}{a_{_0}} \right) \nabla \Phi_N \left( \bm{r} \right) \right] \, , \nonumber
\end{eqnarray}
where $\widetilde{\nu} \equiv \nu - 1$ and $\nu$ is the QUMOND interpolating function, defined such that $\mu \nu = 1$ in spherical symmetry. In this case, $\bm{g} = \nu \bm{g}_{_N}$, with $\nu$ taking the argument $y_\mathrm{int} \equiv \lvert \nabla \Phi_N \rvert/a_{_0}$. This is already much more amenable to computer simulations because $\bm{g}_{_N}$ and any simple function of it are readily calculable using standard techniques. This formulation of MOND also follows from a Lagrangian \citep{Milgrom_2010}, ensuring the usual symmetries and conservation laws with respect to the linear and angular momentum and the energy. For consistency with Equation~\ref{mu_cases}, $\nu$ must satisfy the following asymptotic limits:
\begin{eqnarray}
	\nu \left( y_\mathrm{int} \right) \to \begin{cases}
	\frac{1}{\sqrt{y_\mathrm{int}}} \, , & \textrm{if} ~g_{_N} \ll a_{_0} \left( y_\mathrm{int} \ll 1 \right) \, , \\
	1 \, , & \textrm{if} ~g_{_N} \gg a_{_0} \left( y_\mathrm{int} \gg 1 \right) \, .
	\end{cases}
	\label{nu_cases}
\end{eqnarray}
\textsc{por} applies the simple interpolating function \citep{Famaey_Binney_2005} because it agrees with recent observations on galaxy scales \citep{Iocco_Bertone_2015, Lelli_2017, Banik_2018_Centauri, Chae_2018, Zhu_2023}.\footnote{A somewhat sharper transition at very high accelerations is needed to match Solar System constraints, especially Cassini radio tracking data from Saturn \citep{Hees_2014, Hees_2016}.}
\begin{eqnarray}
    \nu ~=~ \frac{1}{2} + \sqrt{\frac{1}{4} + \frac{1}{y_\mathrm{int}}} ~=~ \frac{1}{2} + \sqrt{\frac{1}{4} + \frac{a_{_0}}{g_{_N}}} \, .
    \label{nu_simple}
\end{eqnarray}

QUMOND can be solved for the total potential $\Phi$ using already existing Poisson solvers. This entails solving the Poisson equation twice, first to get $\Phi_N$ and its derivatives in order to compute the source term on the right hand side of Equation~\ref{eq:QUMOND}, and then to solve this equation. This method makes a direct comparison between the Newtonian and Milgromian cases easier \citep[for further information about the technicalities of the code, see][]{Wittenburg_2020}. A user manual for \textsc{por} has recently been written \citep{Nagesh_2021} and contains a summary of numerical MOND simulations that used it to explore isolated and interacting disc galaxies.\footnote{The setup of Milgromian disc galaxies is discussed in more detail in \citet{Banik_2020_M33}.} Some of its interesting recent applications have been to the asymmetric tidal tails of open star clusters in the Solar neighbourhood \citep{Kroupa_2022}, the tidal stability of dwarf galaxies in the Fornax Cluster \citep{Asencio_2022}, the formation of the Local Group satellite planes from a past flyby encounter between the Milky Way and M31 galaxies \citep{Bilek_2018, Banik_2022_SP}, and the star formation rates and bar statistics of isolated disc galaxies initialized with a realistic size for their stellar mass \citep{Nagesh_2023}.

In order to perform cosmological simulations with \textsc{por}, we had to make several changes based on the description in \citet{Candlish_2016}. The main difference to the non-cosmological application of \textsc{por} is that to account for the cosmic expansion, we solve the Poisson equation in super-comoving coordinates \citep[][; Appendix~\ref{appendix:super_co-moving_coords}]{Martel_1998, Teyssier_2002}:
\begin{eqnarray}
    \label{eq:5}
    \widetilde{\nabla}^2 \widetilde{\Phi} \left( \bm{\widetilde{r}} \right) ~&=&~ \frac{3}{2}\Omega_m a_e \left[ \frac{\rho \left( \bm{\widetilde{r}} \right)}{\bar{\rho}} - 1 \right] \\ \nonumber
    &+&~ \widetilde{\nabla} \cdot \left[\widetilde{\nu} \left( \frac{ \lvert \nabla \Phi_N \rvert}{a_{_0}}\right)\widetilde{\nabla}\widetilde{\Phi}_N \left( \bm{\widetilde{r}} \right)\right] \, , 
\end{eqnarray}
with $a_e \equiv 1/\left( 1 + z_e \right)$ being the cosmic expansion factor at redshift $z_e$ and a tilde denoting super-comoving coordinates except for $\widetilde{\nu}$. The parameter $\Omega_m \equiv \bar{\rho}/\rho_\mathrm{crit}$ is the fraction that the mean matter density $\bar{\rho}$ currently comprises of the cosmic critical density
\begin{eqnarray}
    \rho_\mathrm{crit} ~\equiv~ \frac{3H_0^2}{8 \mathrm{\pi} G} \, .
    \label{rho_crit}
\end{eqnarray}
An important detail in a cosmological context is that the MOND Poisson equation is only applied to the density contrast relative to the average. This originates from the works of \citet{Sanders_2001} and \citet{Nusser_2002}, which introduced the basic principle behind the growth of density perturbations in a Milgromian framework. It has recently been argued that our approach is valid in some relativistic MOND theories \citep*{Thomas_2023}.

The hydrodynamical evolution of the gas is calculated by solving the Euler equations, which \textsc{por} does using a second-order Godunov scheme. We adopt the simple cooling and heating scheme \citep[][and references therein]{Wittenburg_2020}, which makes use of look-up tables and assumes collisional excitation equilibrium (CIE). The simulations here do not allow stellar particles to form. The more rigorous scheme with radiative transfer and supernovae \citep[see also][]{Wittenburg_2020} will have to be implemented in future higher resolution work. This should be straightforward as it is already coded and merely needs to be activated, though cosmological MOND simulations including star formation would be more computationally intensive. Such simulations have already been done in a non-cosmological context and give rather promising results \citep{Renaud_2016, Wittenburg_2020, Eappen_2022, Nagesh_2023}, with the latter work clarifying what sub-grid parameters would be appropriate in a MOND context and also that the particular choice does not much influence the results.

\subsection{Initial conditions}
\label{sec:initcond}

The setup of a cosmological simulation involves two main steps:
\begin{enumerate}
    \item Specifying a power spectrum for the density fluctuations at the desired starting redshift; and
    \item Sampling this power spectrum using a random number generator to specify the initial density and velocity field.
\end{enumerate}
To generate the power spectrum at the desired starting redshift, post-inflationary density fluctuations $\delta \left( k \right)$ are determined based on the cosmological parameters, with the wavenumber $k$ being the inverse of the comoving length scale of the perturbation. The ratio between the density fluctuation at some early time just after the inflationary epoch and at some later epoch is known as the transfer function $T \left( k \right)$, which is defined so that
\begin{eqnarray}
    \delta \left( k \right) ~\equiv~ \delta_0 \left( k \right) T \left( k \right) \, ,
    \label{eq:transfer-func}
\end{eqnarray}
where $\delta_0$ refers to the early epoch and $\delta$ to the later epoch. The power spectrum $P \left( k \right) \propto \delta^2 \left( k \right)$ is the Fourier transform of the two-point correlation function \citep[e.g.,][]{Eisenstein_1998}. The transfer function needs to evolve inflation-generated density perturbations through the radiation-dominated era, recombination, and part of the matter-dominated era up to the epoch when we wish to start the simulation. To limit the computational cost, linear perturbation theory is usually applied for as long as possible, so an important issue is when the structures become non-linear. In the SMoC, this occurs when $\lvert \delta \rvert \ll 1$ is no longer a good approximation. The linear regime breaks down earlier in $\nu$HDM because the underlying gravity law is non-linear. It was argued in section~3.1.3 of \citet{Haslbauer_2020} that the typical gravitational field enters the MOND regime when $z_e \la 50$, so we start our simulations at $z_e=199$ when typical accelerations are still in the Newtonian regime.

\begin{figure*}
    \includegraphics[width=1.0\linewidth]{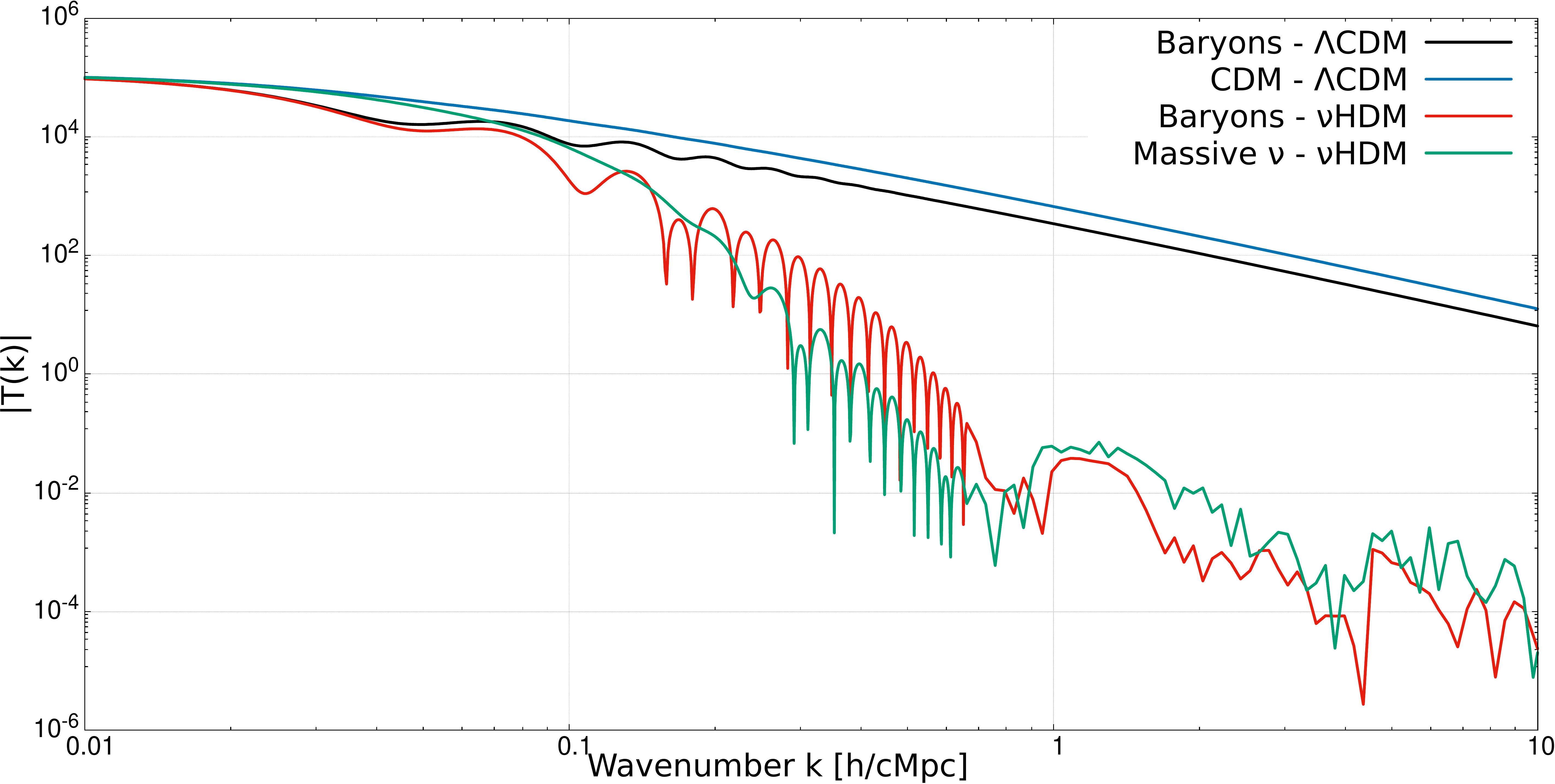}
    \caption{Absolute value of the transfer function $\lvert T \left( k \right)\rvert$ (Equation~\ref{eq:transfer-func}) of the different matter components in $\Lambda$CDM and $\nu$HDM at $z_e=199$, found using \textsc{camb} \citep*{Lewis_2000}. The black (red) line shows the baryons in $\Lambda$CDM ($\nu$HDM), while the blue (green) line shows the CDM in $\Lambda$CDM (neutrinos in $\nu$HDM). Notice the lack of power on small scales in $\nu$HDM due to neutrino free streaming. For the $\nu$HDM components, the transfer function behaves like a damped harmonic oscillator, so the dips to very low $\log \lvert T \left( k \right)\rvert$ correspond to $k$ where the power vanishes.}
    \label{fig:transfer}
\end{figure*}

The transfer function depends on various physical processes like the M\'{e}sz\'{a}ros effect \citep{Meszaros_1974}, acoustic oscillations, Silk damping \citep{Silk_1968}, free streaming, and radiation drag \citep[][and references therein]{Peacock_2003}. In a classical CDM approach, there are no oscillations in the transfer function of the CDM component because it is collisionless and pressureless. The oscillations in the baryonic component stem from the coupling with radiation in the very early universe. After recombination, these oscillations are strongly damped as baryons are forced to follow the pattern of density fluctuations in the dominant CDM component, i.e., the baryons follow the gravitational potential of the CDM (black and blue lines in Figure~\ref{fig:transfer}). In contrast, oscillations are evident for both the baryons and the massive neutrinos in the $\nu$HDM model (green and red lines in this figure). This is expected for the baryons because there is no CDM and thus little power on small scales at $z_e=199$. Since the massive neutrinos have very little power on small scales due to free streaming effects, oscillations in the baryonic power spectrum are able to induce similar oscillations in the sterile neutrino component due to gravitational interactions between the baryons and neutrinos.

The presence of acoustic oscillations in the dominant mass component of the universe as late as $z_e=199$ is a novel feature of the $\nu$HDM paradigm. One important consequence is that the overall transfer function can become negative, indicating that an initial overdensity becomes an underdensity and vice versa. In the linear regime, a purely gravitational problem cannot produce a negative transfer function. This arises in our model and in $\Lambda$CDM at higher redshifts due to the coupling with radiation, which causes the oscillatory behaviour evident in Figure~\ref{fig:transfer}. Note, however, that the sign of $T \left( k \right)$ does not change the power spectrum or the initial density distribution for our simulations because the initial phase of a density fluctuation is random, so only $\lvert T \left( k \right) \rvert$ is relevant for our purposes.

The impact of MOND on the early universe was already discussed in \citet{McGaugh_1999} and references therein. Although the results were based on a purely baryonic model, we still expect the conclusions to be valid regarding MOND producing a more rapid growth of structure and pronounced baryon acoustic oscillations (BAOs) in the transfer function at epochs much later than recombination. Additionally, we expect to detect a difference in the absorption signal of the 21~cm spin-flip transition of neutral hydrogen seen against the CMB. \citet{McGaugh_2018} compared the 21~cm absorption feature in a purely baryonic universe to the $\Lambda$CDM prediction and concluded that the signal in a universe devoid of CDM should be significantly deeper. When comparing $\nu$HDM and $\Lambda$CDM, this discrepancy will certainly be less pronounced, but in general we expect a difference in the 21~cm absorption signal between these two models due to differences in the formation of galaxies and other ionizing sources. In particular, significant BAOs should be evident in the matter power spectrum because non-linear MOND effects (which only become important at $z_e \la 50$) would presumably take some time to wash out the strong BAOs through coupling between perturbations on different scales.

The transfer function for our simulations is calculated using the \textsc{camb} program \citep{Lewis_2000}. We adopt the following parameters: $\Omega_m = 0.314595$, $\Omega_b=0.0492713$, $\Omega_c = 0$, $\Omega_\Lambda = 0.685405$, $\sigma_8 = 0.8101$, $n_\mathrm{spec}=0.9660499$, $H_0 = 67.4 \, \mathrm{km s^{-1}Mpc^{-1}}$ or $h = 0.674$ \citep[this is consistent with the standard parameters from][]{Planck_2020}, and $\Omega_\nu = \Omega_m - \Omega_b = 0.2653237$ for the neutrino component. The gas temperature at $z_e = 199$ is $T=370$~K, which follows from the CMB conditions. The main difference is that we close the gap between $\Omega_b$ and $\Omega_m$ using $11$~eV/$\mathrm{c}^2$ sterile neutrinos instead of CDM particles. The sterile neutrino mass does not need to be specified in \textsc{camb}, but it does display the calculated mass on screen assuming thermalization in the early universe. To implement the sterile neutrinos, we follow the description in \citet{Angus_2009}\footnote{Details are given in section 2.3 of \url{https://cosmologist.info/notes/CAMB.pdf}}, which leads to the addition/alteration of the following lines of the \textsc{camb} ``.ini'' namelist file:
\begin{verbatim}
    massless_neutrinos = 2.0293
    nu_mass_eigenstates = 2
    massive_neutrinos = 1 1
    nu_mass_degeneracies = 1.0147 1
    nu_mass_fractions = 0.0044 0.9956
    accurate_massive_neutrino_transfers = T
\end{verbatim}

\begin{figure*}
    \includegraphics[width=0.495\linewidth]{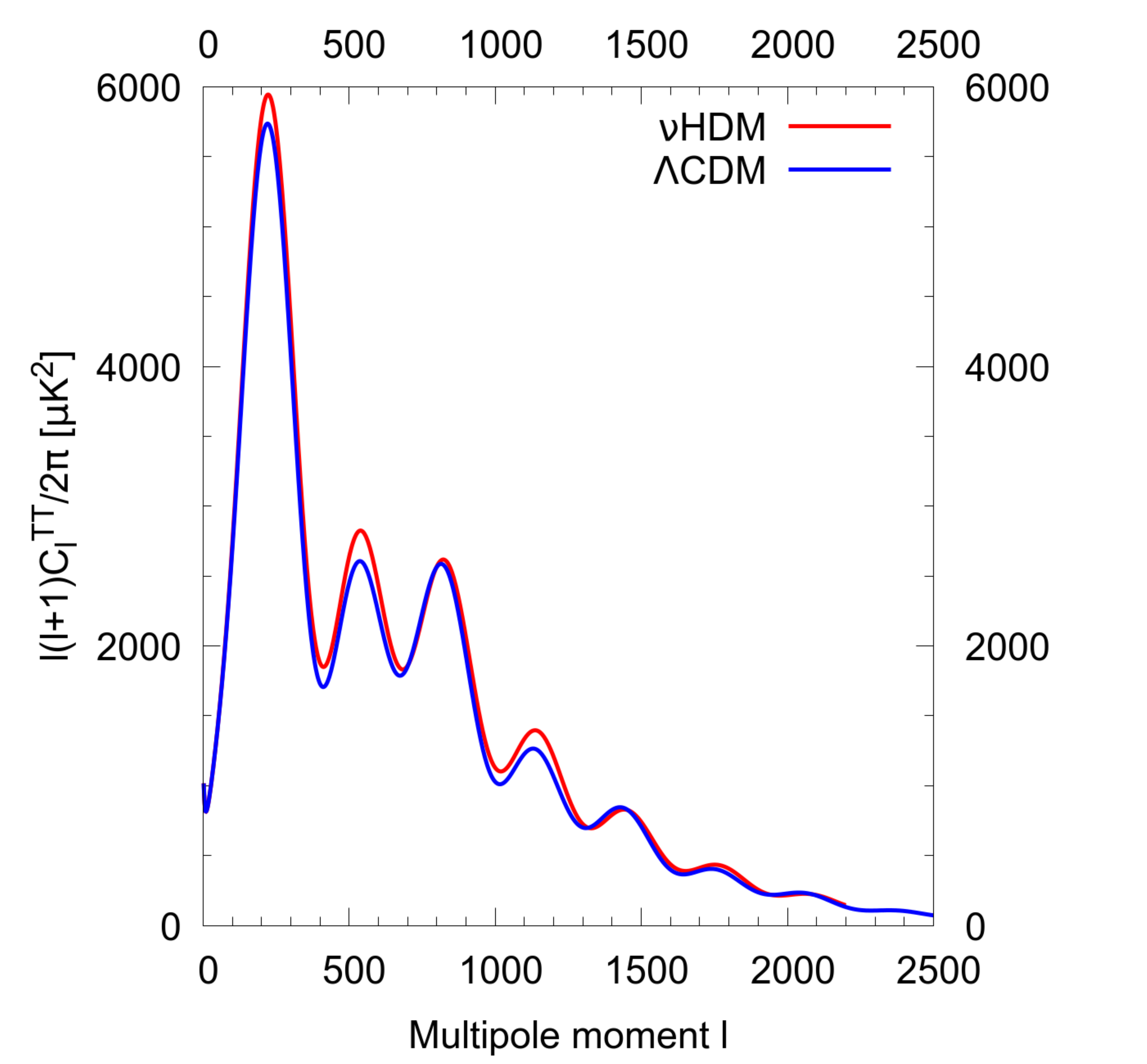}
    \includegraphics[width=0.495\linewidth]{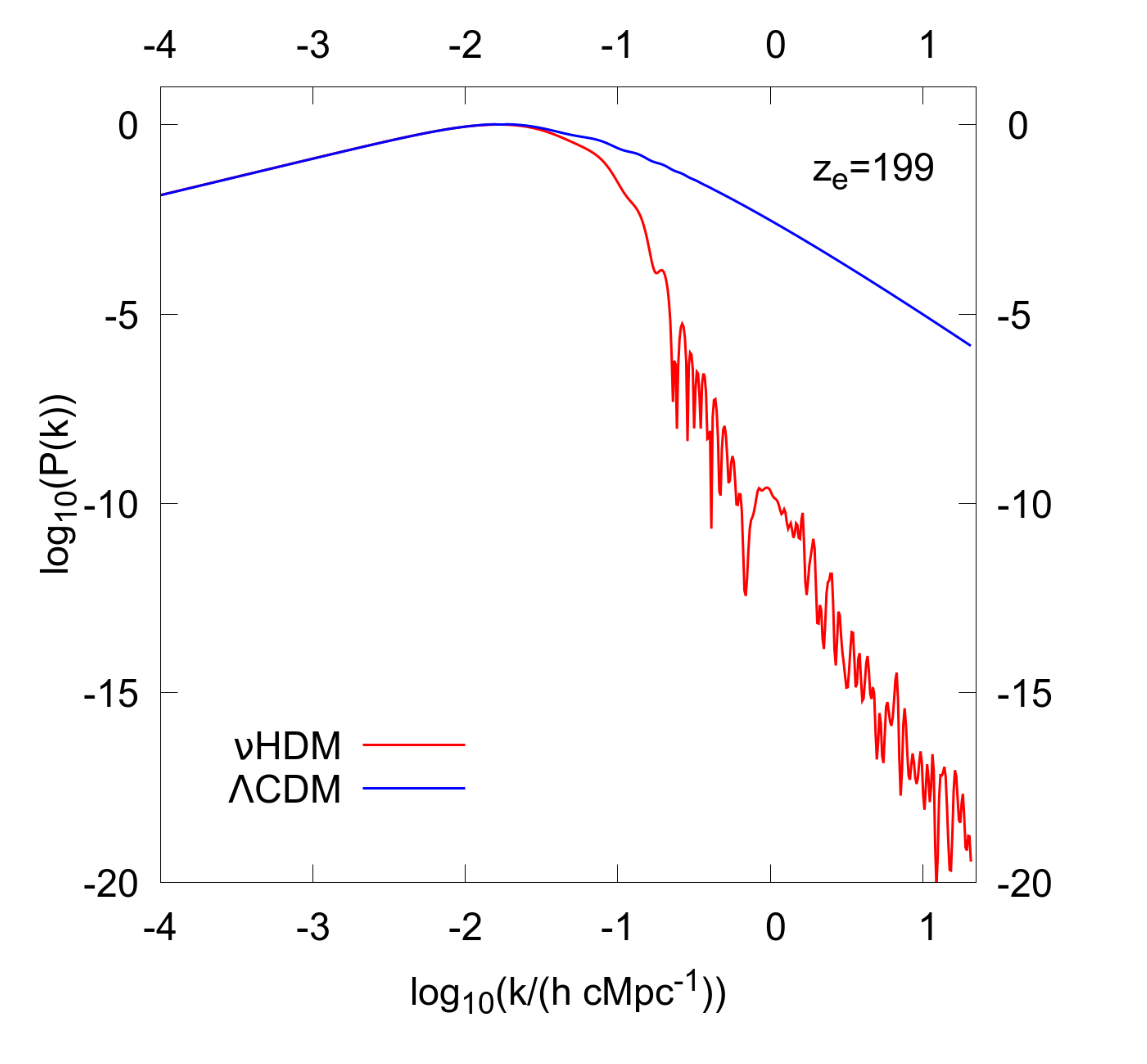}
    \caption{Temperature power spectrum of the CMB (\emph{left}) and total matter power spectrum at $z_e=199$ (\emph{right}) for the $\Lambda$CDM model with standard parameters \citep[blue;][]{Planck_2020} and the $\nu$HDM model considered here (red; see the text). These results were computed with the \textsc{camb} code \citep{Lewis_2000}.}
    \label{fig:powerspec}
\end{figure*}

The left panel of Figure~\ref{fig:powerspec} shows the temperature fluctuations as CMB multipole moments in $\Lambda$CDM with standard cosmological parameters \citep{Planck_2020} and in the $\nu$HDM model simulated here. The $\nu$HDM model fits the observations quite well \citep[as was already shown by][]{Angus_2009}. Further refinements to the cosmological parameters are possible, but we do not consider this here. Figure~1 of \citet{Angus_Diaferio_2011} suggests that finding an optimal fit to the CMB in $\nu$HDM would slightly increase the inferred $H_0$, but not by enough to solve the Hubble tension $-$ we are currently investigating this issue (Luchoomun et al., in preparation).

The right panel of Figure~\ref{fig:powerspec} shows the matter power spectrum in terms of wavenumber $k$ for the $\Lambda$CDM and $\nu$HDM models at our starting redshift of $z_e=199$. The models are practically identical for large scales (small $k$), but the power of the $\nu$HDM model decreases significantly at scales below 100~cMpc. This behaviour is to be expected as the sterile neutrinos do not cluster on small scales by construction. If they did, then the gravitational field on galactic scales would be influenced by them, but there is generally excellent agreement between the MOND dynamical mass of a galaxy and its baryonic mass (Section~\ref{Sec:Introduction}). Another interesting feature of the $\nu$HDM matter power spectrum on small scales is that it shows significant oscillations, which are also evident in the transfer function (Figure~\ref{fig:transfer}). We conclude that these oscillations are driven by the well-known acoustic oscillations in the baryons, which have comparable or more power than the sterile neutrinos on these small scales and can therefore influence them significantly through gravity. The $\Lambda$CDM power spectrum does not show the same oscillations on small scales because the CDM is dominant on these scales and dampens the BAOs, which are a relic of the era when matter was tightly coupled to radiation.

Having specified the power spectrum, the density field is obtained using the publicly available code \textsc{music} \citep{Hahn_2011} but with an externally supplied transfer function (see above), bypassing its inbuilt calculator. \textsc{music} works with the logarithm of the transfer function, which it implicitly assumes is positive. Since this assumption is violated in $\nu$HDM, we adjust the algorithm to work with $\lvert T \left( k \right) \rvert$, which as explained above does not alter the physics. We use three different resolution settings in which $l_\mathrm{max} = l_\mathrm{min} + 1$ in all cases and $l_\mathrm{min}=7$, 8, or 9. This is done in order to prevent the appearance of artificial overdensities, since the number and mass of the sterile neutrino particles is determined by $l_\mathrm{min}$: the total number of sterile neutrino particles is $N_\nu=2^{3 \, l_\mathrm{min}}$ in 3D and the mass of a single sterile neutrino particle is $m_\nu=\Omega_\nu\rho_\mathrm{crit}b^3/N_\nu$, with $b$ being the box size of the simulation volume. We also consider different comoving box sizes of $\mathrm{b}200 \equiv 200$~cMpc/$h$ and $\mathrm{b}400 \equiv 400$~cMpc/$h$, leading to a maximum spatial resolution of 195.3~ckpc/$h$ (b200l10), 390.6~ckpc/$h$ (b200l9 and b400l10), 781.2~ckpc/$h$ (b200l8 and b400l9), and 1562.4~ckpc/$h$ (b400l8), with the number at the end of each simulation code name referring to its maximum refinement level. \textsc{music} needs to generate initial conditions at all the refinement levels which may be used by the main \textsc{por} simulation.\footnote{The cosmological parameters, box size, and resolution settings must be specified consistently between \textsc{camb}, \textsc{music}, and \textsc{por}.}

\subsection{Finding gravitationally bound structures}
\label{Sec:halofinder}

We find gravitationally bound structures using the \textsc{amiga halo finder} \citep[\textsc{ahf};][]{Gill_2004, Knollmann_2009}, partly because it comes with a converter for \textsc{ramses} data, i.e., for converting grid cells into ``cell particles'' located at the cell centre with the same mass.\footnote{The cosmology patch for \textsc{por} and the codes \textsc{ahf}, \textsc{music}, and \textsc{camb} can be found here: \url{https://bitbucket.org/SrikanthTN/bonnpor/src/master/Cosmo_patch_and_setup_and_halofinder/}} Like \textsc{ramses}, \textsc{ahf} is an AMR-based code that uses the adaptive refinement strategy to identify subhaloes at a modest computational cost. 

\textsc{ahf} identifies galaxies depending on the density in a certain volume and afterwards neglects gravitationally unbound structures. However, the criterion for identifying these is not valid in MOND, so we have to use the raw data and skip the latter step. As there is no halo finder currently available that also has the option to use MOND gravity and building one is beyond the scope of this project, we stress that this introduces a bias towards less massive objects. However, the reader should keep in mind that, due to the greater self-gravity, Milgromian structures are generally more stable than Newtonian ones, so many of the identified low mass structures are expected to be gravitationally bound. Additionally, we restrict our analysis to structures that contain baryons because low- and intermediate-mass structures and substructures that only contain sterile neutrinos are most likely chance alignments. Our structures are identified considering both baryons and neutrinos and must have at least 10 cell particles and/or sterile neutrinos. Our analysis of the simulations is still at a rather preliminary stage due to the difficulties created by the use of a non-linear gravity law for which relatively little prior work is available, especially on cosmological scales.

To compare our results with observations, we consider the edge of a structure to be at the radius $R_{180}$, the radius inside which the mean mass density is $180 \, \rho_\mathrm{crit}$ (Equation~\ref{rho_crit}).
\begin{eqnarray}
    \frac{M_{180}}{\left(4/3\right)\mathrm{\pi}R_{180}^3} ~\equiv~ 180 \, \rho_\mathrm{crit} \, ,
    \label{M_180}
\end{eqnarray}
with the mass considering both baryons and sterile neutrinos. The halo finder calculates the mass and radius iteratively starting from an initial guess.

There is an additional problem when comparing observations with MOND simulations: for a given gravitational field, the corresponding mass profile depends on the gravity law. Since mass is not a direct observable, the mass deduced in the context of an assumed gravity law is called the dynamical mass in that theory. Most works in the literature quote the Newtonian dynamical mass profile $M_N \left( r \right)$, which is the enclosed Newtonian dynamical mass within radius $r$. By definition, $M_N \left( r \right) \equiv r^2 g \left( r \right)/G$ for a spherical system, with the gravitational field $g \left( r \right)$ generally being what the observations actually constrain through kinematical information. In MOND, the Newtonian gravitational acceleration is enhanced by a factor $\nu$, but this enhancement can also be achieved within Newtonian gravity if we rescale the enclosed mass $M \left( r \right)$ by this factor \citep{Angus_2011}:
\begin{eqnarray}
    M_N \left( r \right) ~=~ \nu \left( g_{_N} \left( r \right) \right) M \left( r \right) \, .
    \label{eq:boost}
\end{eqnarray}
We include Equation~\ref{eq:boost} in the halo finder, so every $M_{180}$ value reported here is actually the Newtonian dynamical mass obtained from the total mass (gas + neutrinos) in the simulation data by applying the above boost \citep[this issue is discussed further in section~4.3 of][]{Asencio_2021}. This allows for an easier comparison with observations as the literature usually gives the Newtonian dynamical mass rather than the MOND one.

\subsection{Simulation parameters}
\label{Sec:models}

Every simulation done for this work starts at $z_e = 199$ or $a_e = 0.005$ because the typical gravitational field from inhomogeneities enters the MOND regime only when $z_e \la 50$ \citep[see section~3.1.3 of][]{Haslbauer_2020}, so our choice achieves a good compromise between limiting the effect of MOND at later times and limiting the impact of radiation, which is dominant only at much earlier times. We run six models: b200l10, b400l10, b200l9, b400l9, b200l8, and b400l8. The labels ``10'', ``9'', and ``8'' stand for the maximum refinement level used in the simulation (see Section~\ref{sec:initcond}), while ``200'' and ``400'' show the length of the simulation box in cMpc/$h$. These models are summarized in Table~\ref{table:galtot}.

\begin{table}
    \centering
    \small
    \caption{Global properties of all simulations at the present time. $N_{\mathrm{tot,struc}}$ is the total number of structures in the simulation box, $M_{\mathrm{struc,bar}}$ is the total baryonic mass in all structures, $M_{\mathrm{tot,bar}}$ is the total gas mass (which remains constant throughout the simulation), $N_\nu$ is the total number of sterile neutrinos in the simulation box, and $m_\nu$ is the mass of a sterile neutrino particle in units of $10^{10} \, M_\odot$ (see Section~\ref{sec:initcond}). $f_{\mathrm{{struc,bar}}} \equiv M_{\mathrm{struc,bar}}/M_{\mathrm{tot,bar}}$ is the fraction of the baryonic mass in structures. Masses are given in units of $10^{17} \, M_\odot$ and the maximum spatial resolution (res\textsubscript{max}) in ckpc/$h$. Note that $M_{\mathrm{struc,bar}}$ decreases with increasing maximum resolution res\textsubscript{max}, which is expected to some degree as these simulations are not numerically converged and coarser resolution can lead to an overestimation of the mass inside structures.}
    \begin{tabular}{ccccccc} 
     \hline
        Model & b200 & b400 & b200 & b400 & b200 & b400 \\
        & l10 & l10 & l9 & l9 & l8 & l8 \\ \hline
        $N_{\mathrm{{tot,struc}}}$ & 518 & 1509 & 416 & 1467 & 217 & 687 \\ 
        $M_{\mathrm{{struc,bar}}}$ & 0.66 & 3.84 & 0.69 & 4.42 & 0.67 & 4.58 \\
        $M_{\mathrm{{tot,bar}}}$ & 1.62 & 12.98 & 1.62 & 12.98 & 1.62 & 12.98 \\
        $f_{\mathrm{{struc,bar}}}$ & 0.41 & 0.30 & 0.43 & 0.34 & 0.41 & 0.35 \\
        res\textsubscript{max} & 195.3 & 390.6 & 390.6 & 781.2 & 781.2 & 1562.4 \\
        $N_{\nu}/10^6$ & 134.2 & 134.2 & 16.8 & 16.8 & 2.1 & 2.1 \\
        $m_{\nu}$ & 0.65 & 5.21 & 5.21 & 41.68 & 41.68 & 333.42 \\
     \hline
    \end{tabular}
    \label{table:galtot}
\end{table}

All simulations were performed on in-house SPODYR-group machines in Bonn, which have 64 CPU cores, 128 threads, and 512 GB of RAM. The high-resolution run (model b200l10) took a computational time of 5 days with 64 threads, corresponding to 7680 CPU hours.

\section{Results}
\label{Sec:results}

Our simulations do not contain a key feature of standard simulations, namely CDM particles creating deep potential wells into which baryons fall after cooling \citep[e.g.,][]{White_1978, Frenk_2012}. Galaxies are nevertheless generally expected to form earlier in MOND than in the SMoC \citep{Sanders_1998_cosmology, Sanders_2001} because of the strong long-range enhancement to the gravitational acceleration. Resolving the formation of small structures is highly dependent on the maximum refinement level and the box size. Due to the lack of power on small scales (see the right panel of Figure~\ref{fig:powerspec}) and the influence of structures on each other through tides and the external field effect \citep[EFE; see sections 2.4 and 3.3 of][]{Banik_Zhao_2022}, it is crucial to simulate large enough boxes in order to properly capture structure formation in the $\nu$HDM model. This makes it very taxing to do high resolution simulations, especially hydrodynamical ones.

In the highest resolution model (b200l10), the first structures can only be identified when $z_e=4$. This late formation of structures seems to be in discordance with observations: there are many very massive galaxies at $z_e \ga 4$, a fact which has become clearer with deeper and improved observations \citep{Steinhardt_2016, Marrone_2018, Wang_2019, Rennehan_2020, Neeleman_2020, Tsukui_2021}, including very recently with data from the James Webb Space Telescope \citep[JWST; e.g.,][]{Furtak2022, Harikane2022, Labbe2022, Naidu2022a, Haslbauer_JWST_2022, Yan2023, Adams2023, Artek2023}. We discuss this issue in more detail in Section~\ref{Sec:Time_evolution}.

\begin{figure*}
    \includegraphics[width=0.495\linewidth]{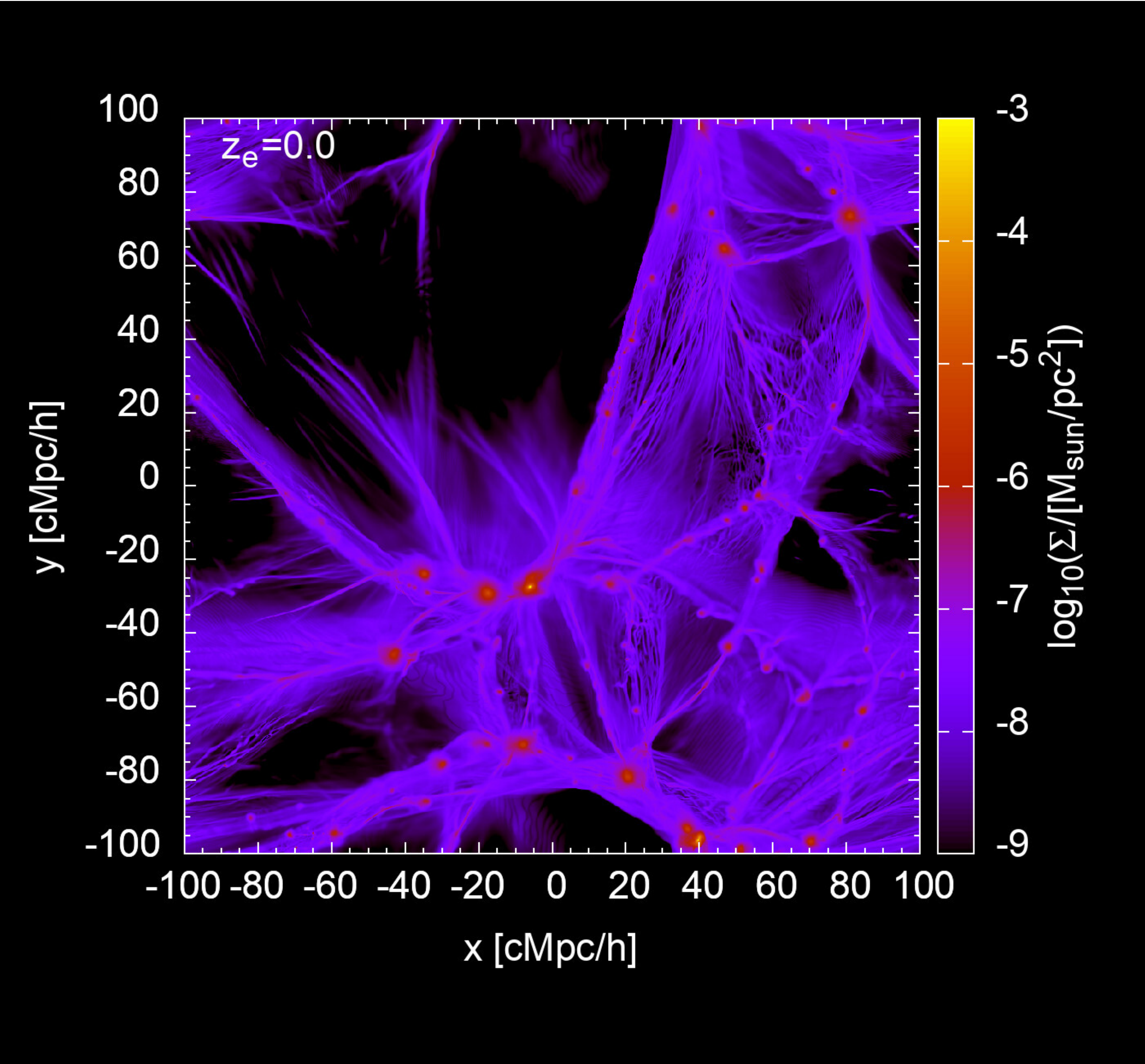}
    \includegraphics[width=0.495\linewidth]{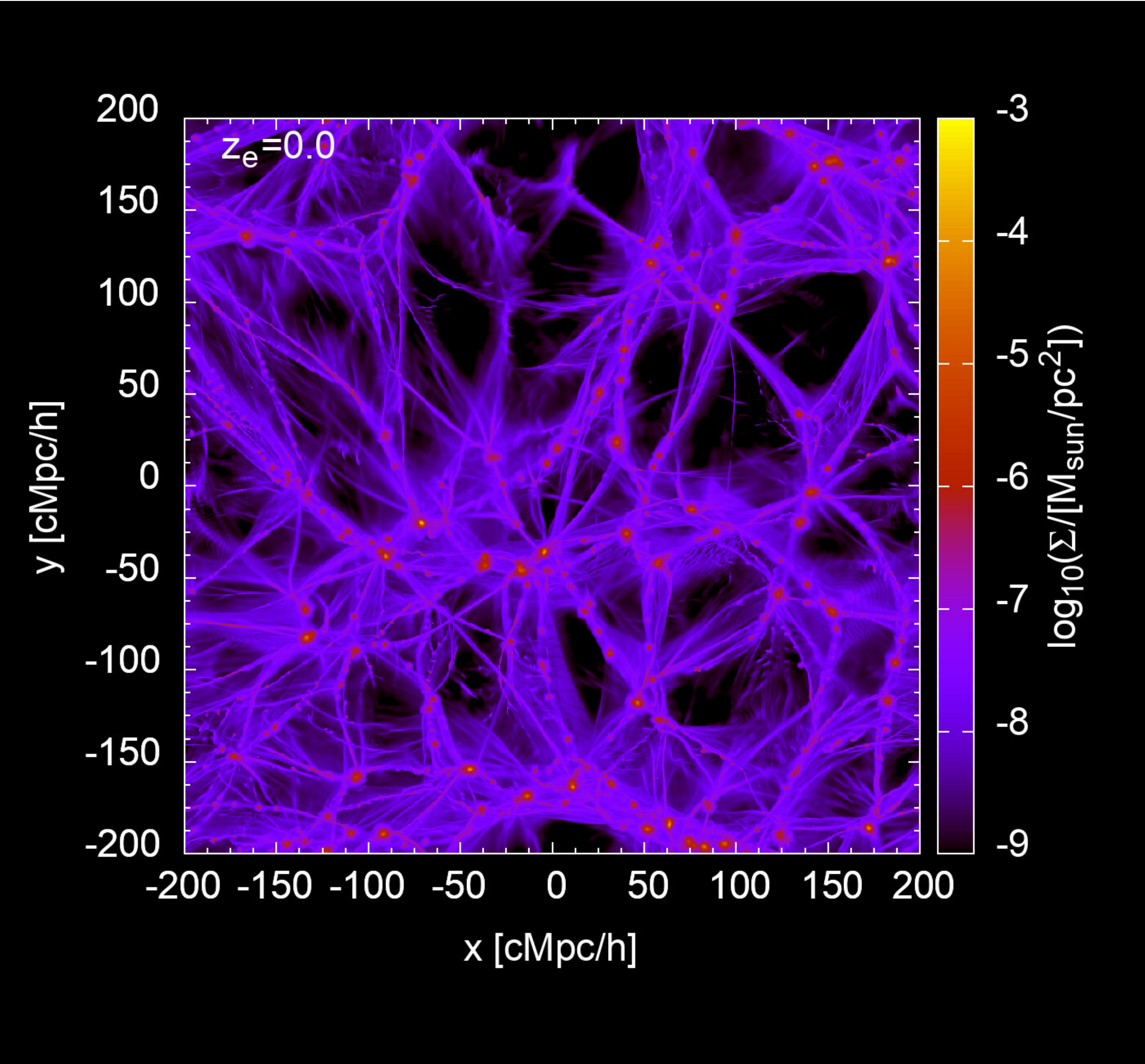}
    \includegraphics[width=0.495\linewidth]{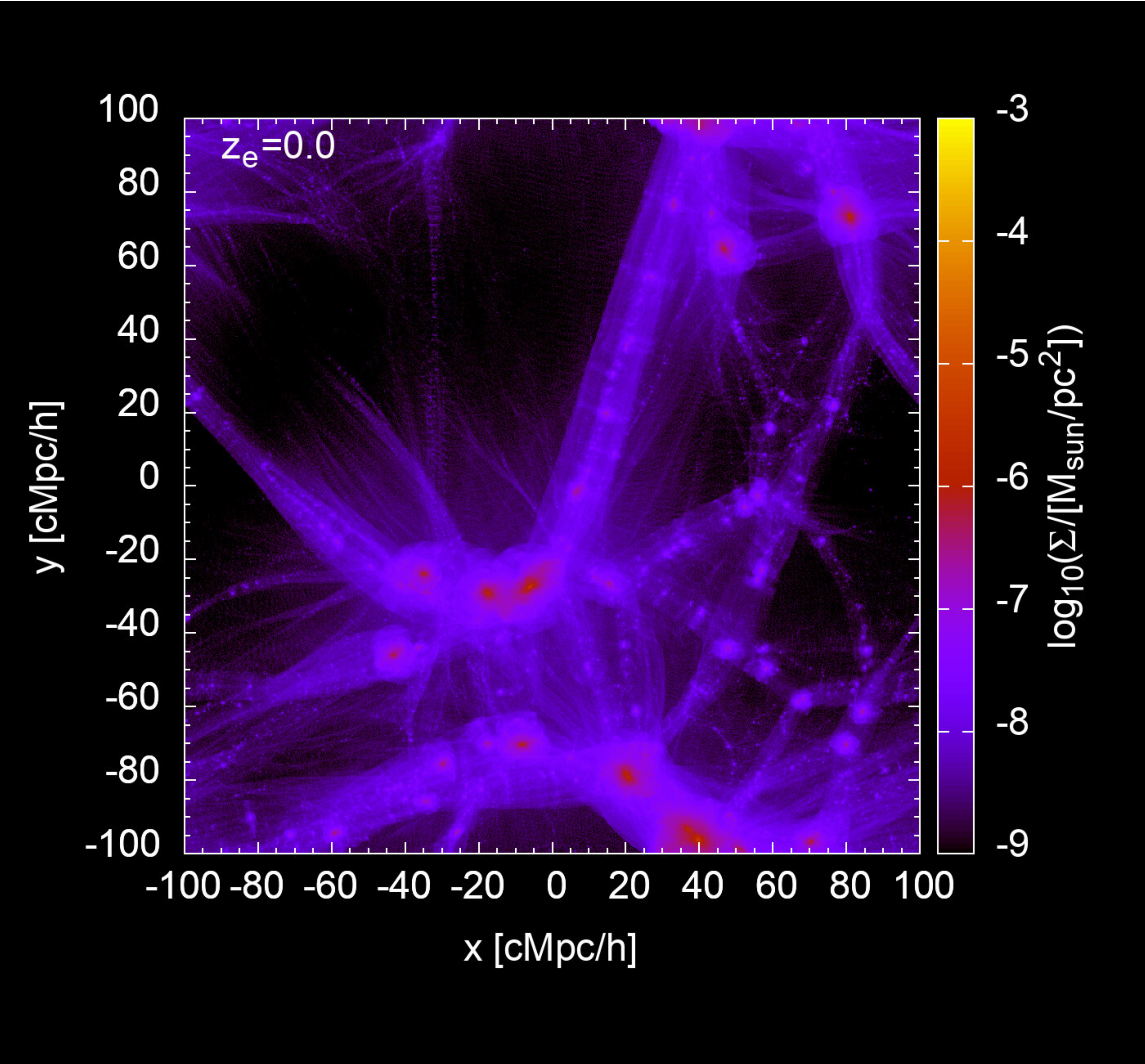}
    \includegraphics[width=0.495\linewidth]{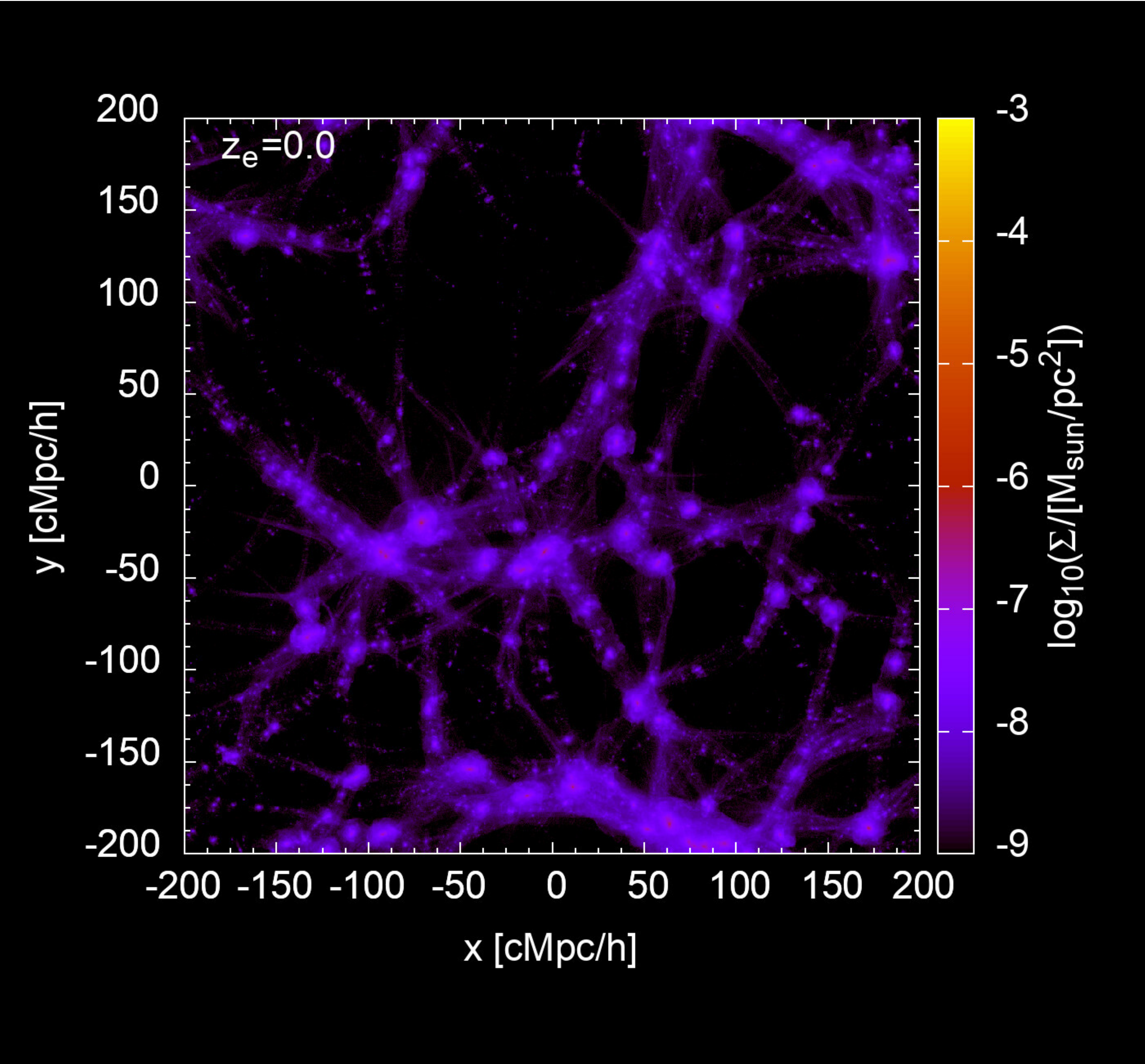}
    \caption{\textit{Top panels}: Baryonic surface mass density (colour-coding) in the $xy$ plane of the two highest resolution models (left panel: b200l10, right panel: b400l10) at a redhsift of $z_e = 0$. The surface mass density shown here is calculated by integrating the mass-weighted 3D mass density along the projection axis for each pixel. \textit{Bottom panels}: Surface mass density for the sterile neutrinos, which is calculated by computing the total projected mass in a pixel and dividing it by the area of the pixel. Movies of the simulations are available at \url{https://www.youtube.com/playlist?list=PL2mtDSIH4RQhtoIxlOBvDzrzhT7bAFxqq}.}
    \label{fig:densityz0}
\end{figure*}

Figure~\ref{fig:densityz0} shows the structures which form in our models by the present epoch, with the top panels showing the baryons and the bottom panels showing the sterile neutrinos.\footnote{Movies of the growth of structure can be found here: \url{https://www.youtube.com/playlist?list=PL2mtDSIH4RQhtoIxlOBvDzrzhT7bAFxqq}.} Comparing the left and right panels of Figure~\ref{fig:densityz0} illustrates how the box size affects the growth of structure throughout cosmic time: model b400l10 (right panels) has significantly more structure than b200l10 (left panels). This demonstrates the difficulty of achieving ``Copernican convergence'', i.e., having a sufficiently large volume such that the statistical properties do not change if a larger volume is considered. Copernican convergence is harder to achieve in MOND because the gravity law is non-linear, so the effect of a structure in the simulation on other structures declines much more slowly with distance. Moreover, structures on Mpc scales form in MOND in a top-down manner because there is very little power on small scales initially. Our models are not fully Copernican converged due to computational limitations.

\subsection{The mass function}
\label{sec:mass-function}

\begin{figure*}
    \includegraphics[width=1.0\linewidth]{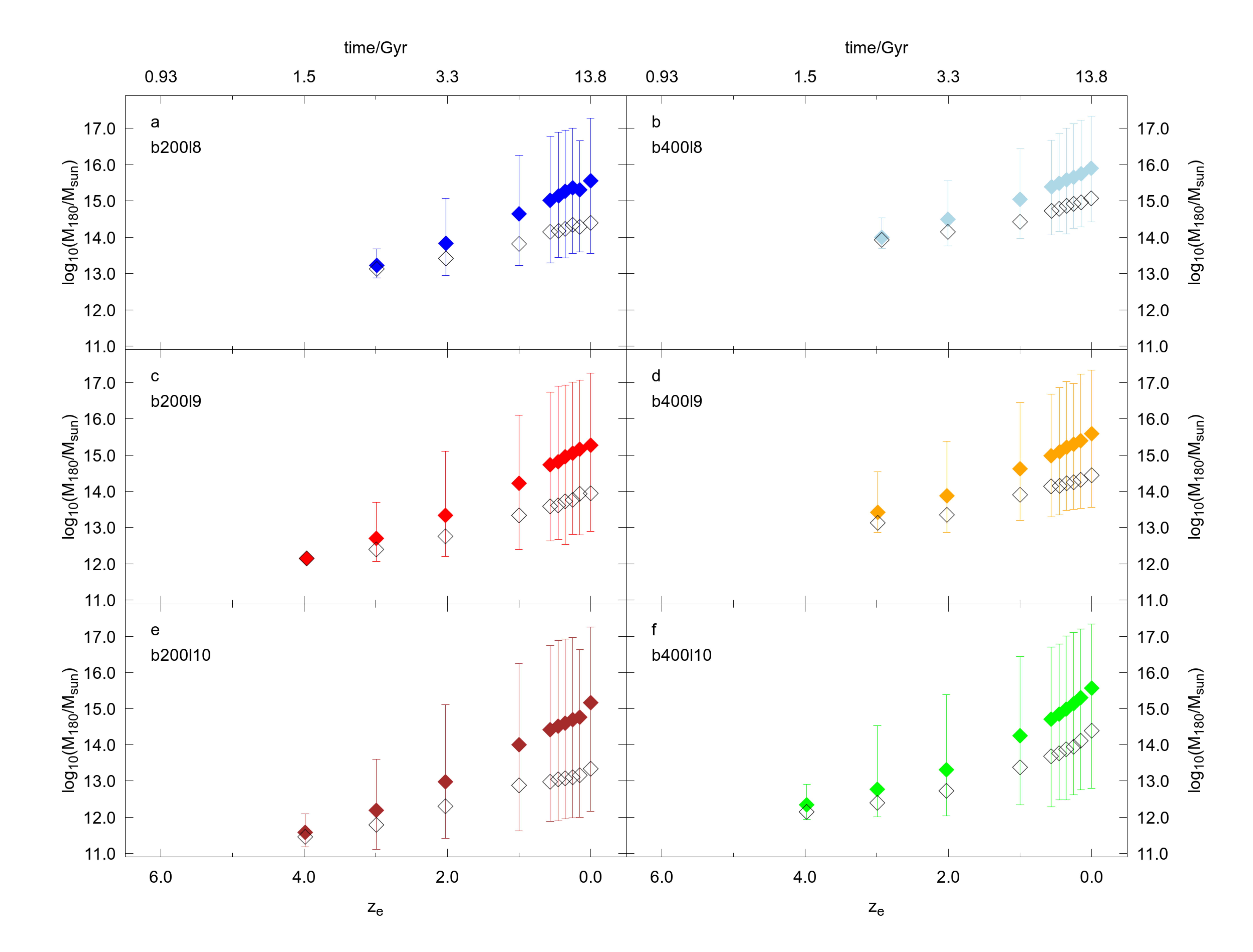}
    \caption{Distribution of the structure masses (baryons + neutrinos) at every output for every model. The filled (open) diamonds show the mean (median) mass of all structures at each epoch, while the error bars show the full range. The left (right) panels show the models with a box size of 200 (400) cMpc/$h$. The resolution increases from top to bottom, with the $l_\mathrm{max}=8$ models at the top and the $l_\mathrm{max}=10$ models at the bottom.}
    \label{fig:mass-time}
\end{figure*}

As our first quantitative analysis, we discuss the mass distribution of the formed structures and how this evolves with time. Figure~\ref{fig:mass-time} shows the mass distribution of structures in all six models at different times (left: b200 models, right: b400 models, top: $l_\mathrm{max}=8$, middle: $l_\mathrm{max}=9$, bottom: $l_\mathrm{max}=10$). The mass of the most massive structure (top of each error bar) and the least massive structure (bottom of each error bar) grow over time, as is also the case for the mean and median mass (the filled points and black open points, respectively). The impact of resolution is apparent in that the first structures are identified earlier in the higher resolution models. The mass of the most massive structure increases when we consider a larger box, which is expected because larger structures can arise in a larger volume. These preliminary results are thus by no means numerically converged. However, they do clearly illustrate the impact of box size and resolution. For instance, the least massive structures in the b400 models (right panels) are more massive than the least massive structures in the b200 models with the same $l_\mathrm{max}$ (left panels) because the spatial resolution is higher in the latter case.

\begin{figure*}
    \includegraphics[width=1.0\linewidth]{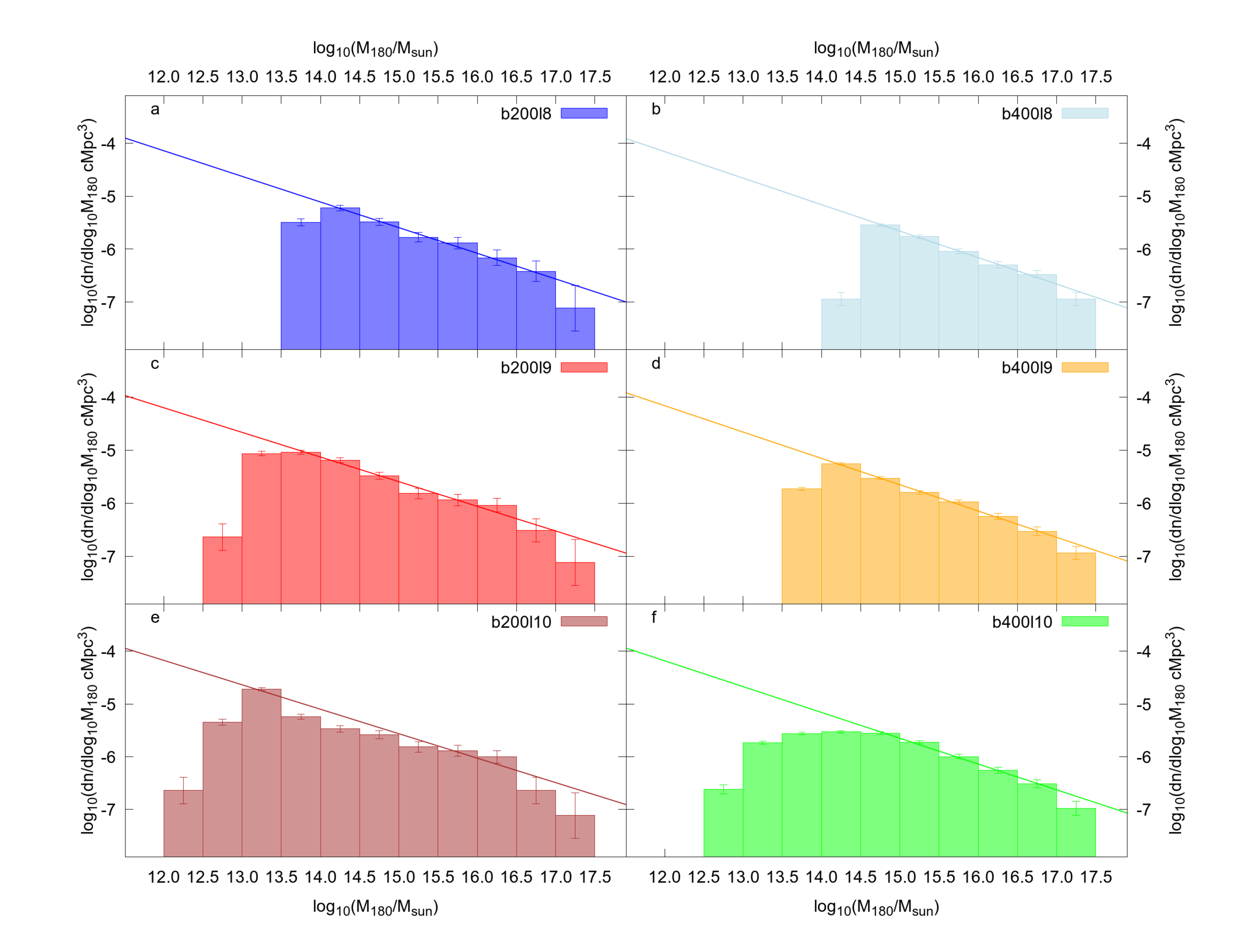}
    \caption{The differential mass function at $z_e=0$ for every model. The left (right) panels show the models with a box size of $b=200$~cMpc/$h$ ($b=400$~cMpc/$h$). The resolution increases from top to bottom, with the $l_\mathrm{max}=8$ models at the top and the $l_\mathrm{max}=10$ models at the bottom. The mass function was calculated by dividing the number of structures per bin by the product of the total simulation volume and the width of the mass bin in dex (this being 0.5). The error bars show Poisson uncertainties and the solid lines show the power-law fits to the mass function (Equation~\ref{eq:powerlaw} and Table~\ref{table:fit_cluster}).}
    \label{fig:massfunction}
\end{figure*}

Figure~\ref{fig:massfunction} shows the present-day ($z_e=0$) differential mass function (baryons + neutrinos) of structures in all our simulations, with the panels organized in the same way as in Figure~\ref{fig:mass-time}. We place the structures into mass bins of width 0.5~dex and divide the number counts by the product of the bin width and the total comoving volume under consideration. The distributions are then fit by a power law:
\begin{eqnarray}
    \frac{dn}{d\log_{10} M_{180}} ~\equiv~ \xi_{struc}\left( M_{180} \right) ~=~ A\left( \frac{M_{180}}{M_\odot}\right)^{-\alpha}\, ,
    \label{eq:powerlaw}
\end{eqnarray}
where $A$ is the normalization and $\alpha$ is the power-law index. The fitted values for these quantities are shown in Table~\ref{table:fit_cluster} for every model.

\begin{table}
    \centering
    \caption{Parameters of the power-law (Equation~\ref{eq:powerlaw}) fits shown in Figure~\ref{fig:massfunction}. The two fit parameters are not independent, so the uncertainty on the normalization in the range covered by the data is much smaller than the quoted uncertainty on $A$ (this is especially apparent in model b400l9). Since the typical mass is of order $10^{15} \, M_\odot$, we expect the uncertainty on $\log_{10} A$ to be about $15\times$ that on $\alpha$, which is roughly the case.}
    \begin{tabular}{ccc} 
    \hline
    Model & $\log_{10} A$ & $\alpha$ \\ \hline
    b200l8 & 1.68 $\pm$ 0.44 & 0.48 $\pm$ 0.03 \\ 
    b400l8 & 1.85 $\pm$ 0.34 & 0.50 $\pm$ 0.03 \\
    b200l9 & 1.39 $\pm$ 0.46 & 0.47 $\pm$ 0.04 \\
    b400l9 & 1.81 $\pm$ 0.40 & 0.50 $\pm$ 0.03 \\
    b200l10 & 1.39 $\pm$ 0.98 & 0.46 $\pm$ 0.07 \\
    b400l10 & 1.69 $\pm$ 0.52 & 0.49 $\pm$ 0.04 \\ \hline
    \end{tabular}
    \label{table:fit_cluster}
\end{table}

Figure~\ref{fig:massfunction} illustrates the mass distribution of the structures at the end of the simulations. When comparing simulations of the same box size (in the same column), the effect of resolution is clearly evident in that the lowest mass bin gets more populated at higher resolution. Additionally, the shape of the distribution seems to be changing from a simple power law closer to a Schechter function \citep{Schechter_1976} with higher resolution, which is the observationally expected shape $-$ this is especially evident for the b200 models. However, increasing the box size shifts the distribution to more massive structures. Not only are the most massive structures found in the larger box (as is evident from Figure~\ref{fig:mass-time}), but also the number density of such structures is higher in a larger box.

\begin{figure*}
    \centering
    \includegraphics[width=0.495\linewidth]{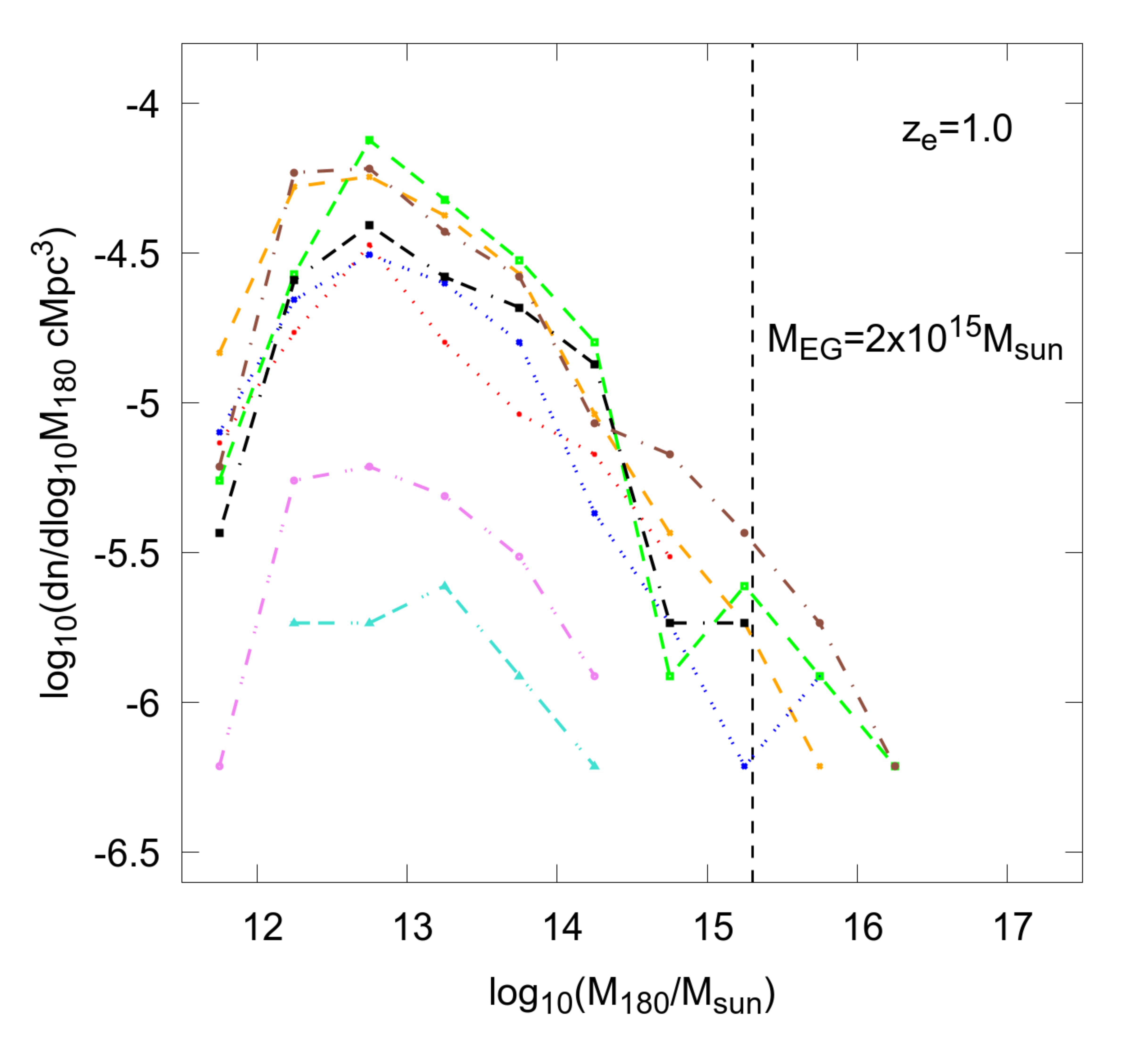}
    \includegraphics[width=0.495\linewidth]{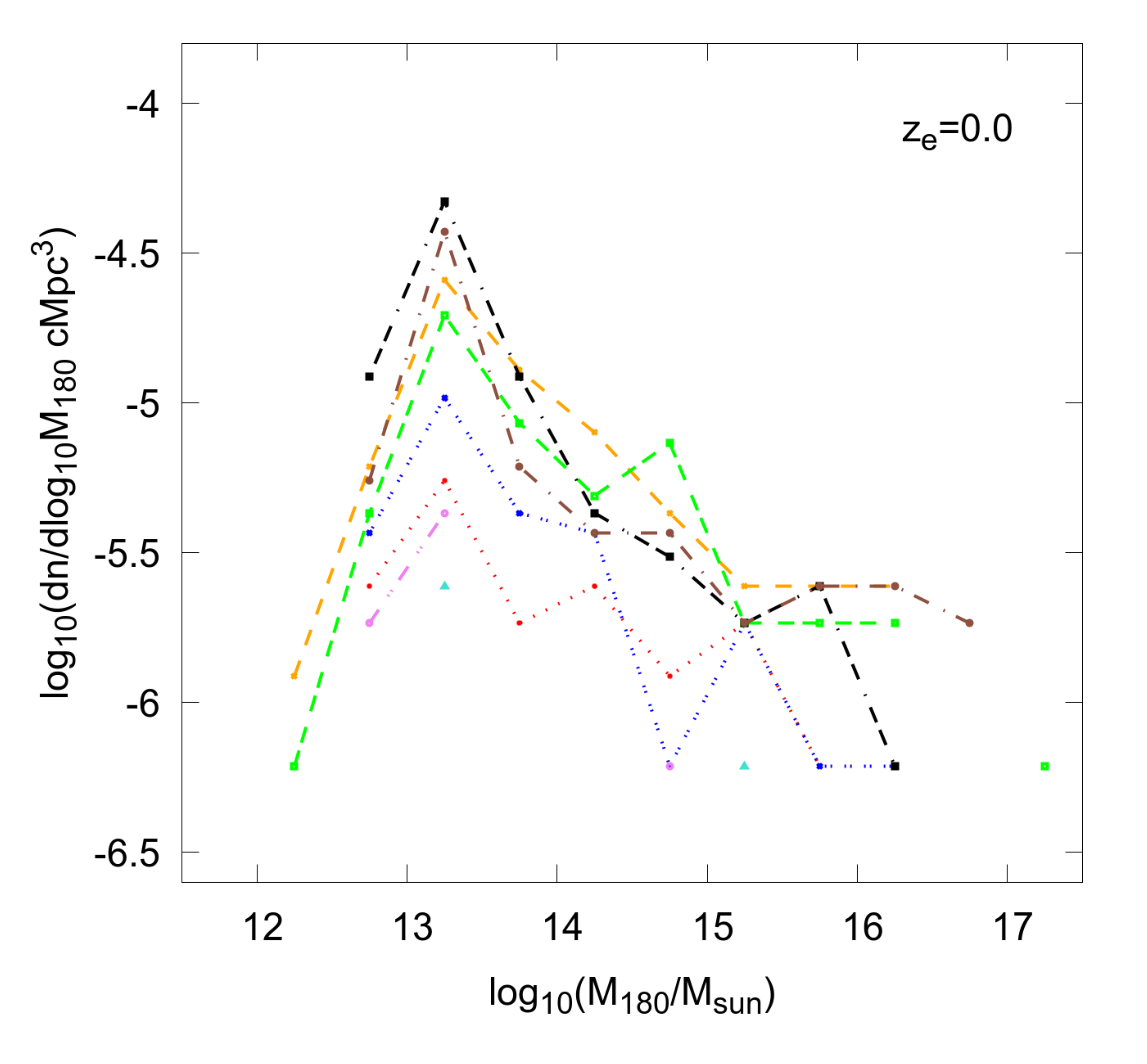}
    \caption{Mass functions calculated similarly to Figure~\ref{fig:massfunction}, but now showing only the model b200l10 divided into 8 equally sized cubes at two different redshifts of $z_e = 1.0$ (\textit{left panel}) and $z_e=0.0$ (\textit{right panel}). Every line thus corresponds to a simulation cube with box size $b=100$~cMpc/$h$, which is half of the initial box size. The dashed vertical line in the left panel shows the mass of the El Gordo galaxy cluster \citep[$M_{\mathrm{EG}} \approx 2 \times 10^{15} \, M_\odot$;][]{Kim_2021}.}
    \label{fig:massfunction100Mpc}
\end{figure*}

To estimate the cosmic variance in our simulations, we divide the whole b200 simulation cube into 8 equally sized cubes (octants) with sides of length $b = 100$~cMpc/$h$, which is half that of the original simulation. Figure~\ref{fig:massfunction100Mpc} shows the mass function in each octant for model b200l10 at $z_e = 1.0$ (left panel) and today (right panel). The aforementioned existence of pronounced voids and overdense regions is also evident here in that there is more than one dex difference between the mass function in the least and most dense octants. Additionally, the figure shows that the mass of the most massive structure depends on the environment: the least dense octant forms structures with masses up to $M_{180} = 10^{15.0} \, M_\odot$, while the most dense sub-volumes contain structures with $10^{17.0} \, M_\odot < M_{180} < 10^{17.5} \, M_\odot$ at $z_e=0.0$. Qualitatively the same phenomenon is evident at $z_e=1.0$ but shifted to lower masses, reaching $10^{14.0} \, M_\odot < M_{180} < 10^{14.5} \, M_\odot$ for the least dense octant, while the two densest octants reach masses $M_{180} > 10^{16} \, M_\odot$. 

The $\nu$HDM model seems to explain the existence of very massive galaxy clusters at high redshift like the El Gordo cluster \citep{Menanteau_2012}: at $z_e=1$, the $\nu$HDM mass function extends up to $M_{180} \approx 10^{16} \, M_\odot$. El Gordo has a total mass of about $2 \times 10^{15} \, M_\odot$ \citep{Kim_2021} and is composed of two interacting sub-clusters with a similar mass and an infall velocity of $v_{\mathrm{infall}}=2500$~km~$\mathrm{s}^{-1}$ at a redshift of $z_e=0.87$ \citep{Zhang_2015}. Since El Gordo is observed after pericentre, the two progenitor clusters need to be spatially close at $z_e=1$ and have a combined mass of $\approx 2 \times 10^{15} \, M_\odot$. Given also the limited sky area in which El Gordo was discovered, this is plausible only if the mass function at that time reaches somewhat higher masses $-$ which it does in some octants (indicated by the vertical dashed line in the left panel of Figure~\ref{fig:massfunction100Mpc}). Our results are compatible with the work of \citet{Katz_2013}, who found El Gordo analogues in their $\nu$HDM simulations with box sizes of $\approx750$~cMpc (see their section~5.1). By comparing their result to the effective volume of the discovery survey, it was shown that $\nu$HDM produces approximately the correct frequency of El Gordo analogues \citep[see section~4.3 of][]{Asencio_2021}.

\begin{figure*}
    \centering
    \includegraphics[width=0.84\paperwidth, height=0.7\paperheight]{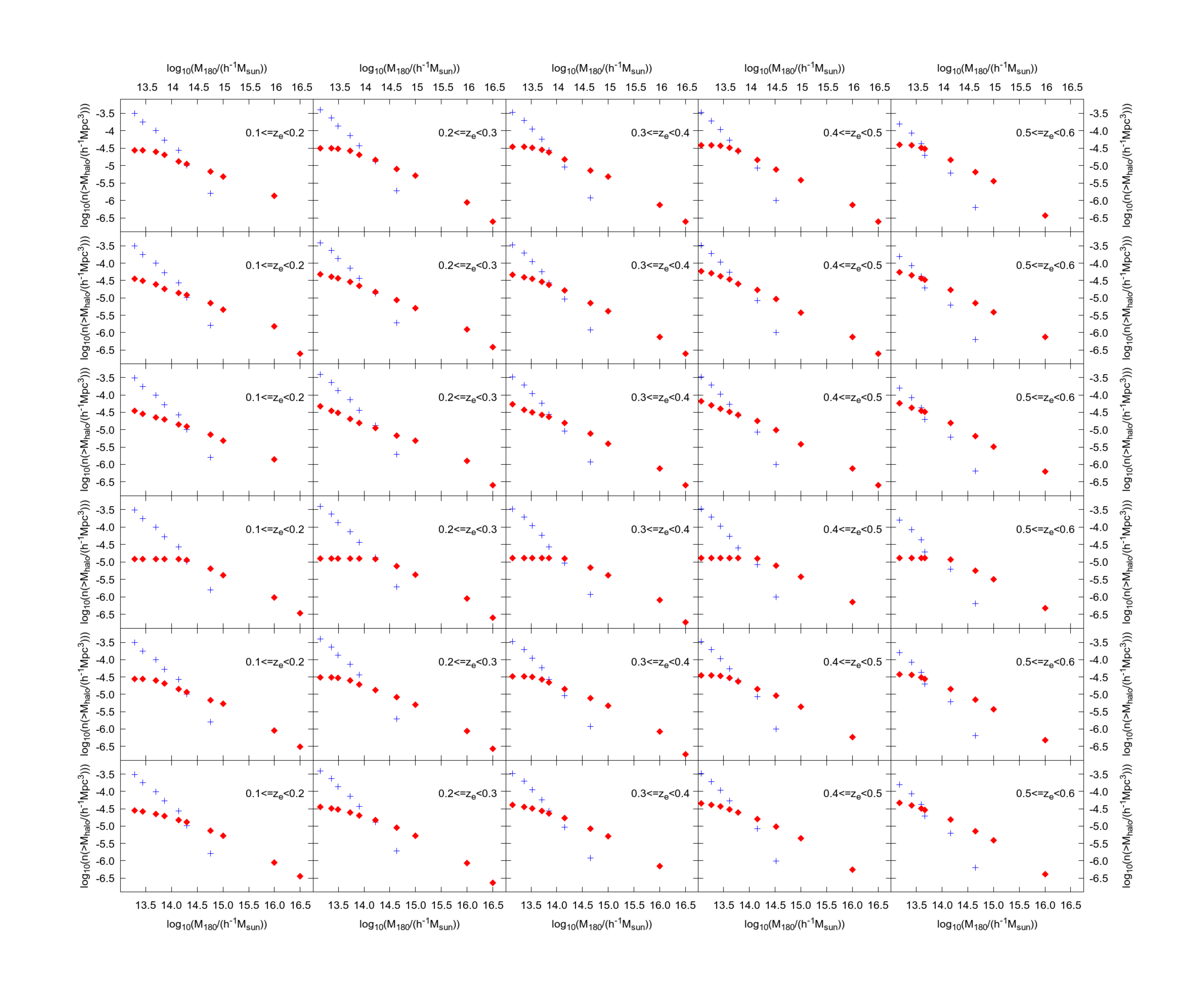}
    \caption{The cumulative mass function of all models compared to observational results based on weak lensing \citep{Wang2022}. The CMF is calculated by counting the number of structures more massive than the mass points taken from their work and dividing by the respective volume for five different redshift ranges. Each row depicts the CMF for one model (from top to bottom: b200l8 to l10, followed by b400l8 to l10), while each column shows the same redshift range (from left to right: $0.1<z_e<0.2$ to $0.5<z_e<0.6$). The red diamonds represent results from simulations, while the blue crosses show observations.}
    \label{fig:cumulative_MF}
\end{figure*}

After analysing in detail the mass distribution of the structures in our simulations, we use Figure~\ref{fig:cumulative_MF} to compare them with the observed mass distribution of galaxy clusters. \citet{Wang2022} obtained weak lensing masses for the clusters from the catalogue of \citet{Yang2021}, which was generated by the halo-based group finder originally developed in \citet{Yang2005}. The observations are based on the Dark Energy Spectroscopic Instrument (DESI) Legacy Imaging Surveys \citep{Dey_2019} and corrected for completeness. The cumulative mass function (CMF) is calculated by counting the number of clusters beyond certain mass thresholds and dividing by the respective volume under consideration for five different redshift slices with a width of $0.1$, thus covering the range $z_e=0.1-0.6$.

For an accurate comparison, we apply the same method to our simulations: we count the number of structures above the mass thresholds used in \citet{Wang2022} and divide by the volume of the respective simulation for outputs that fall into the redshift range, i.e., we compare the observed redhsift bins only with outputs that are in the same redshift range. Figure~\ref{fig:cumulative_MF} shows the CMF for all five redshift bins and all six simulations. Every column depicts the same redshift bin for different models (top to bottom: b200l8, b200l9, b200l10, b400l8, b400l9, and b400l10); while every row shows the redshift evolution for one model, with redshift increasing from left to right.

The most obvious finding is that all models seemingly overproduce high mass structures in all redhsift bins, which we turn to momentarily. The simulated CMFs are also significantly flatter than observed. The two agree approximately at intermediate masses, but there is an obvious impact of resolution on the $\nu$HDM CMFs at the low mass end. The models b200l8, b400l8 and b400l9 show a nearly horizontal CMF at lower masses, indicating that there are no structures at or below these masses (see also Figure~\ref{fig:massfunction}). As anticipated, the number density at any fixed mass increases with decreasing redshift for both observations and our models, which can be attributed to accretion and mergers (cf. Figure~\ref{fig:mass-time}). Taking the discrepancy between observations and our models at high masses at face value puts significant doubt on the theory. A similar conclusion was drawn by \citet{Angus_2013} and \citet{Katz_2013}, who also did cosmological simulations in the $\nu$HDM framework. However, we argue below that their conclusions were premature.

Shortly after the publication of \citet{Angus_2013} and \citet{Katz_2013}, it became apparent that for redshifts up to $z_e=0.07$ (corresponding to a distance of $\approx300$~Mpc), a mismatch in the high mass range is to be expected given our location in a large underdensity known as the KBC void \citep{Keenan_2013}. Their study and the more recent study of \citet{Wong_2022} cover 90\% of the sky, indicating that the observational results on the KBC void are not due to incomplete coverage. They also probe a large portion of the galaxy luminosity function. It is therefore likely that the mass function of observed clusters in the Local Volume is not representative of an average region of the same size. Interestingly, \citet{Haslbauer_2020} examined the KBC void and concluded that it naturally explains why the locally determined Hubble-Lemaitre constant $H_0$ exceeds that determined from CMB anisotropies \citep[the Hubble tension is reviewed in][]{Valentino_2021}. Such a void was initially deemed to be incompatible with the supernova distance-redshift relation \citep{Kenworthy_2019}, but subsequent analyses have found that supernova data alone actually reveal a mild preference for a local void \citep{Lukovic_2020, Kazantzidis_2020}. Indeed, the results of \citet{Castello_2022} indicate that the preferred void parameters are remarkably close to that of the KBC void, highlighting a consistency between results obtained from galaxy number counts and from supernovae. Further, \citet{Dainotti_2021, Dainotti_2022} and \citet{Jia2023} investigate the redshift evolution of $H_0$ and find clear evidence for a smooth decrease from the high local value down to the lower value inferred from the CMB \citep{Planck_2020} out to $z_e=2$. \citet*{Schiavone_2023} interpret these results via an effective Hubble constant that evolves with the redshift in an $f \left( R \right)$ modified gravity scenario, thus indicating an intrinsically different local value of $H_0$. This is very interesting given how \citet{Haslbauer_2020} argued that a local underdensity can change the measured $H_0$: the finding of a smooth transition might indicate that we are situated in a much larger underdensity, which in turn would lead to the local measurements not representing an average region of the Universe. Indeed, figure~4 of \citet{Haslbauer_2020} shows that their inference on the void radius has an extended tail that reaches very large values. Further support for the Universe being more inhomogeneous than assumed by the SMoC comes from \citet{Migkas_2021}, who analysed 570 clusters in the local Universe and built 10 different cluster scaling relations out of these to test the hypothesis of isotropy. They found a dipole-like anisotropy in the local value of $H_0$ with more than $5\sigma$ significance. This study underlines the question raised before: Is the local Universe representative of an average region of the Universe?

One way to test cosmological models despite a possible negative answer is to consider observations of the most massive clusters up to $z_e=1$. El Gordo is not only more massive than the observed clusters closer to us shown in \citet{Wang_2022}, but it is also further in the past. We would expect clusters today to reach an even higher mass. \citet{Asencio_2021} analysed El Gordo in detail by looking at realistic hydrodynamical simulations of the pre-merger configuration and comparing the viable configurations to large-scale $\Lambda$CDM simulations. Those authors found that ``detecting one pair with its mass and redshift rules out $\Lambda$CDM cosmology at 6.16$\sigma$''.

The two sub-clusters that merged into El Gordo need to have masses of about $10^{15}\,M_\odot$ at a redshift close to $z_e=1.0$. From observations out to $z_e=0.6$, it seems unlikely to find even one cluster with such a high mass, but to have two of them in close proximity at a larger redshift strongly indicates that there should be more structure on large scales than is generally expected and observed locally. Encouragingly, \citet{Asencio_2021} argue in detail in their section~4.3 that the previous $\nu$HDM simulations done by \citet{Katz_2013} show the correct frequency of El Gordo analogues. As a result, the apparent overproduction of massive structures \citep{Angus_2013} may well be incorrect as their results were based on comparison with low-redshift datasets. We therefore argue that the $\nu$HDM theory cannot be ruled out by the discrepancies at high masses evident in Figure~\ref{fig:cumulative_MF}. In fact, the $\nu$HDM simulations reported here confirm that bound structures with an El Gordo-like mass readily appear at $z_e=1$ (Figure~\ref{fig:mass-time}) and also show that these mass scales are only reached in dense regions of the simulation box at that redshift (see Figure~\ref{fig:massfunction100Mpc}). Given that the mass of El Gordo is now known to about 10\% accuracy \citep{Kim_2021}, any realistic cosmological model must be able to reproduce its properties, especially given the other potentially problematic clusters and cluster pairs discussed in the introduction to \citet{Asencio_2021}.

\subsection{Time evolution of bound structures}
\label{Sec:Time_evolution}

One of the major processes affecting the galaxy population in the SMoC is mergers between galaxies, which are driven by dynamical friction between extended CDM haloes \citep{Kroupa_2015}. Since these are absent in MOND, the baryonic parts of galaxies would need to overlap for them to merge $-$ they would simply fly past each other in the absence of dissipation \citep{Renaud_2016, Bilek_2018, Banik_2022_SP}. The collision cross-section between galaxies is thus much smaller in MOND, which may explain the high observed fraction of thin disc galaxies \citep{Peebles_2020, Haslbauer_2022}.

\begin{figure*}
    \includegraphics[width=0.49\linewidth]{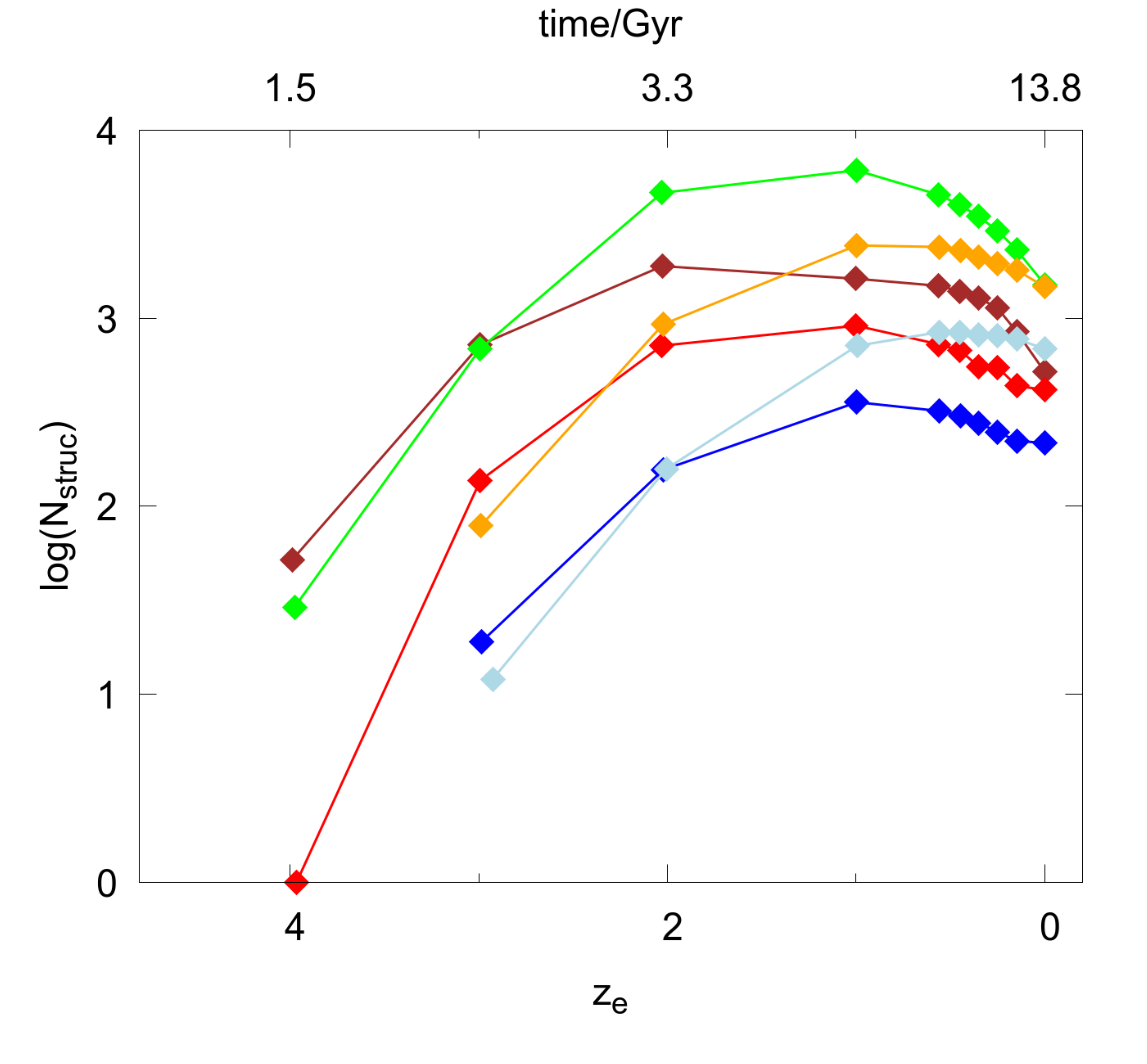}
    \includegraphics[width=0.49\linewidth]{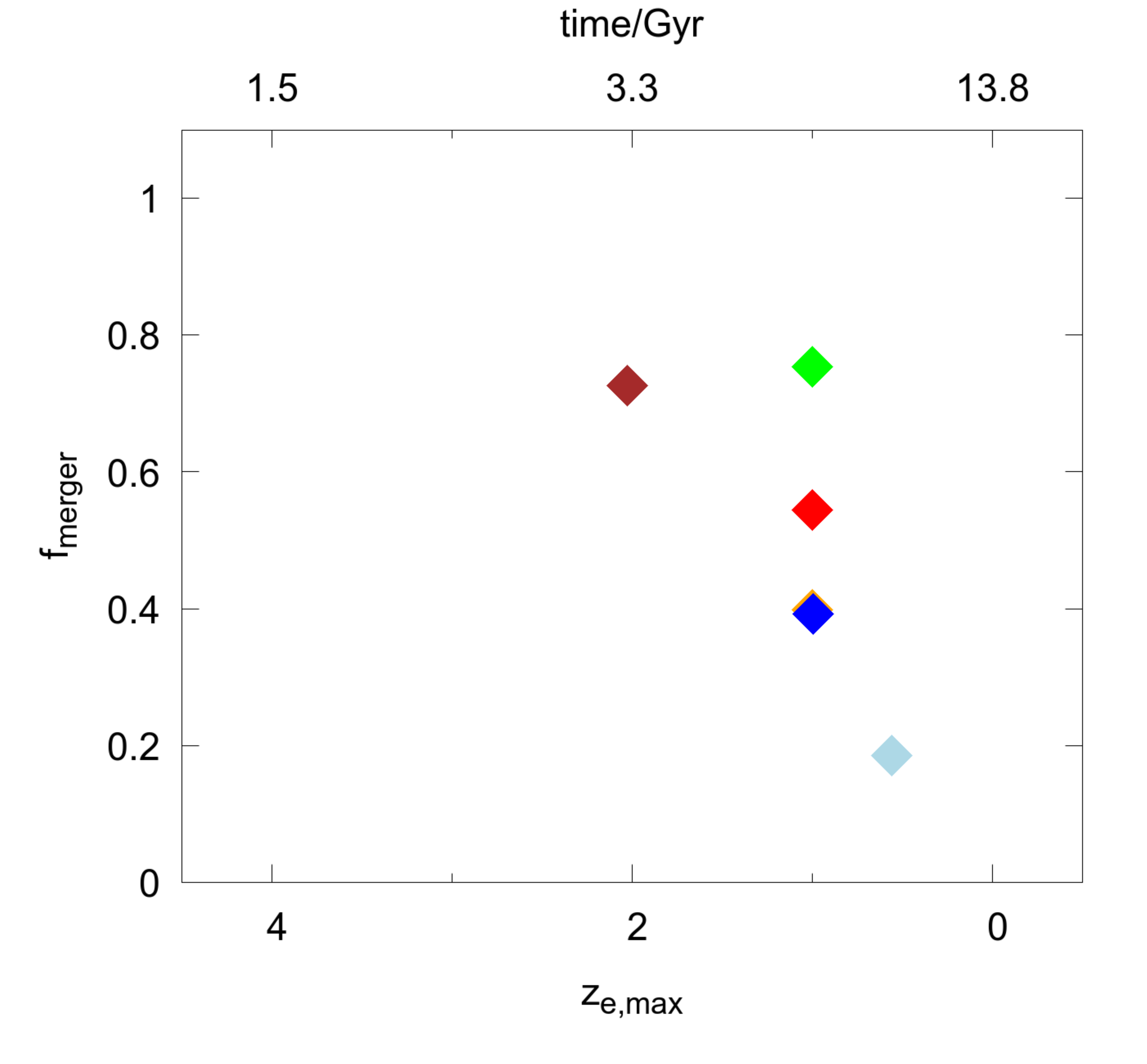}
    \caption{\textit{Left}: Time evolution of the total number of structures for all models. \textit{Right}: The fractional reduction in the number of structures between the epoch at which the number of structures was maximal and the present epoch (Equation~\ref{f_merger}). The brown (b200l10) and green (b400l10) points show the number of structures for the models with the highest resolution at each box size $-$ these also reach the largest $N_\mathrm{struc}$. Colours are as in Figure~\ref{fig:mass-time}. The orange data point (model b400l9) is mostly hidden underneath the dark blue data point (model b200l8).}
    \label{fig:N_time}
\end{figure*}

To explore the time evolution of the population of structures in MOND, we use Figure~\ref{fig:N_time}. Its left panel shows how the total number of structures ($N_{\mathrm{struc}}$) evolves over time for every model, while the right panel shows the fractional reduction in the number of structures between the epoch when this was maximal and the present epoch. In other words, it shows the merger fraction
\begin{eqnarray}
    f_{\textrm{merger}} ~\equiv~ \frac{N_{\textrm{max}} - N_0}{N_{\textrm{max}}} \, ,
    \label{f_merger}
\end{eqnarray}
where $N_{\mathrm{max}}$ is the maximum number of structures (attained at redshift $z_\mathrm{max}$) and $N_0$ is the present number of structures in the simulation volume. This is only an estimated merger fraction because it is possible that new structures form at later epochs in these simulations. However, we expect this to be rather rare because the mass of the least massive structure rises with time (see Figure~\ref{fig:mass-time}).

The evolution of $N_{\mathrm{struc}}$ is qualitatively similar for all models. The total number of structures rises steeply after the onset of structure formation at about $z_e=4$ until it reaches a maximum at $z_e=1$. Only the highest and the lowest resolution model (b200l10, brown and b400l8, light blue) peak earlier and later, respectively. Afterwards, $N_{\mathrm{struc}}$ declines modestly to significantly, depending on the resolution: higher resolution models show a steeper decline. This indicates that mergers play a role in hydrodynamical MONDian models that lack star formation. Important to note is that the observed star formation rate density also peaks at a redshift of $z_e\approx2$ \citep{Madau_2014}, indicating that the number of structures may also peak at this time, which is compatible with our highest resolution model (b200l10). Interestingly, the models with the same box size end with nearly the same number of structures for $l_\mathrm{max}=9$ and 10, which seems to indicate that the last output ($z_e=0$) is much closer to numerical convergence compared to the early universe in these simulations. Additionally, as expected, the simulations with an initially larger box (the b400 models shown in green, orange, and light blue) end up with a larger number of structures.

The right panel of Figure~\ref{fig:N_time} shows the fraction of mergers of structures in each simulation (Equation~\ref{f_merger}). There is a clear dependence on resolution, showing that increasing $l_\mathrm{max}$ and/or decreasing the box size increases $f_\mathrm{merger}$. Also, the maximum of the number of structures mentioned above is more easily visible here, with most of the models peaking at $z_e=1$. As shown in Figure~\ref{fig:mass-time}, the models computed here are not able to reach single galaxy scales at the onset of structure formation (here $z_e\approx4$). \citet{Sanders_1998_cosmology} already estimated the time at which spheres of different masses should collapse in an idealized scenario, i.e., without the EFE and in spherical symmetry. These first analytic calculations estimate the onset of cluster formation at about $z_e=3$, which is consistent with the onset of structure formation in the models shown here. Therefore, higher resolution simulations are needed that at least probe mass scales down to $10^{10} \, M_\odot$ in order to conclusively address if structures form early enough in the $\nu$HDM model compared to the most recent observations. However, it should be noted that the analytical work of \citet{Sanders_1998_cosmology} was not based on the $\nu$HDM framework and in particular has a different background cosmology, i.e., an open universe with $\Omega_b = 0.02$ and no dark matter or dark energy, potentially causing difficulties fitting the CMB and other cosmological datasets collected over the last quarter century. We note that their adopted cosmology gives almost twice as much time for structure formation by $z_e = 10$ because the different expansion history makes the universe older at that redshift. Moreover, increasing the resolution of the $\nu$HDM models causes only a modest evolution towards higher redshift for the onset of structure formation, so it seems unrealistic to form galaxy mass structures at $z_e \geq 10$. This may already indicate a failure of the model.

While these first hydrodynamical cosmological MOND results suggest that the population of structures builds up until $z_e \approx 1 - 2$ and then decreases again due to mergers, the results should be considered indicative rather than conclusive. This is because the merger rate is undoubtedly enhanced by virtue of all baryons being purely gaseous and thus able to dissipate energy rather efficiently during interactions. Future simulations will need to include star formation to obtain a more realistic assessment of the late-time merger rate and also a higher resolution to analyse the behaviour of individual galaxies.

\begin{figure*}
    \includegraphics[width=1.0\linewidth]{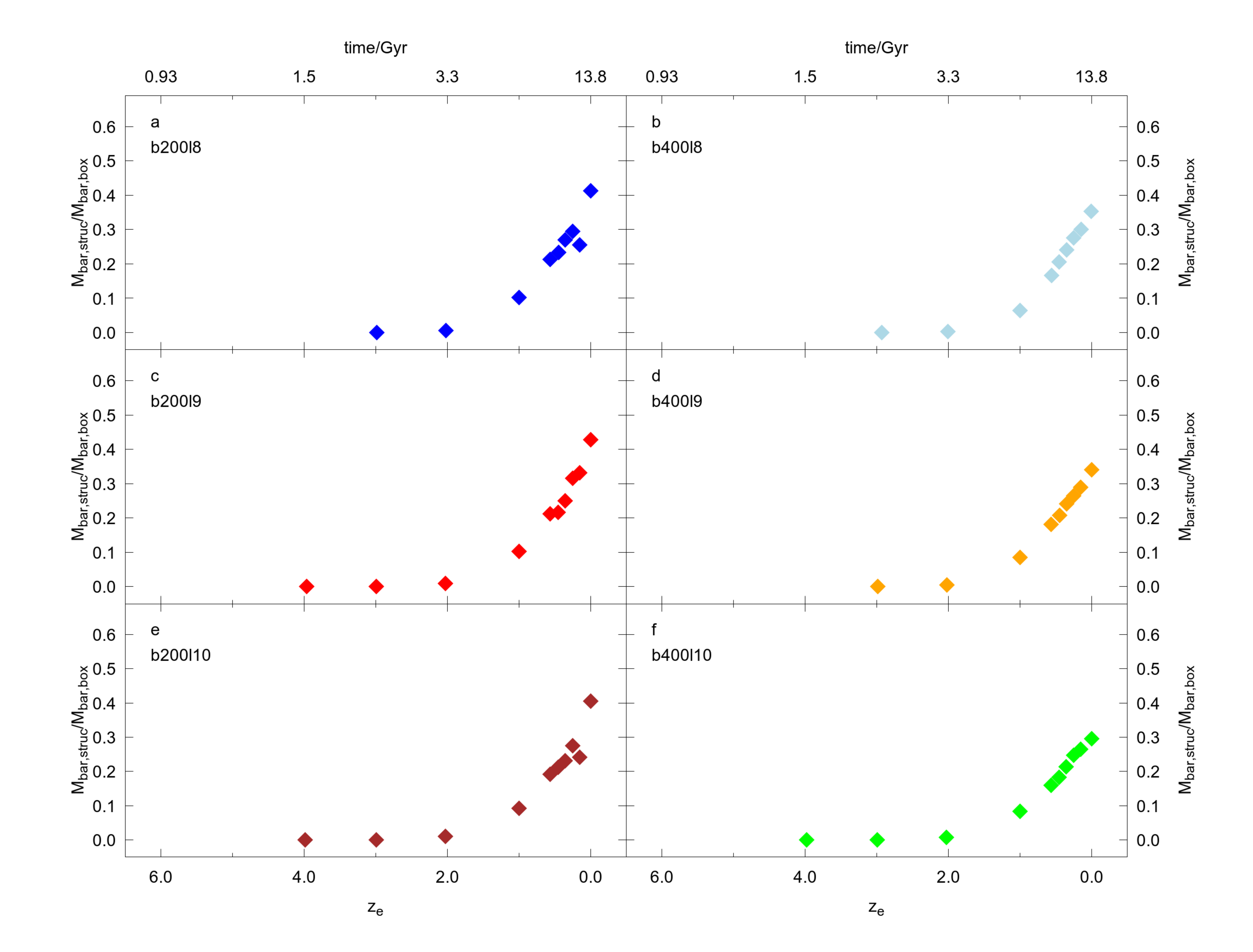}
    \caption{Time evolution of the baryonic mass fraction inside structures. $M_{\mathrm{bar,struc}}$ is the total baryonic mass inside all structures and $M_{\mathrm{bar,box}}$ is the baryonic mass inside the box, which does not change with time (see also Table~\ref{table:galtot}). The symbols and models in each panel are the same as in Figure~\ref{fig:mass-time}.}
    \label{fig:mfraction}
\end{figure*}

Another aspect of the simulations which changes with time is the total baryonic mass inside structures. Its time evolution is shown in Figure~\ref{fig:mfraction}, where $M_{\mathrm{bar,struc}}$ is the total baryonic mass inside all structures and $M_{\mathrm{bar,box}}$ is the baryonic mass inside the box, which does not change with time. A distinct feature at $z_e=2$ can also be seen here: at later times, the baryonic mass fraction increases significantly in the same way for all models up to a value of $0.3-0.5$. This behaviour of all models indicates that the increase for $z_e<2$ is driven by the accretion of baryons on to large structures rather than the formation of new structures or mergers thereof, which is further supported by the fact that $f_\mathrm{merger}$ is significantly lower for the low resolution models where small structures cannot form at high $z_e$. These results indicate that massive enough regions have collapsed and formed deep potential wells at $z_e\approx 2$, such that significant amounts of baryonic mass move through filaments of the cosmic web towards these structures, where the filaments intersect and clusters form \citep[this is compatible with the theoretical findings of][]{Sanders_1998_cosmology}. What is more, a modest trend with box size is evident for the present epoch baryonic mass fraction in identifiable structures, which is closer to $0.4$ for the b200 models and closer to $0.3$ for the b400 models. Additionally, there is also a minor decrease with increasing resolution regardless of the box size.

\subsection{The sterile neutrino mass fraction}

\begin{figure*}
    \includegraphics[width=1.0\linewidth]{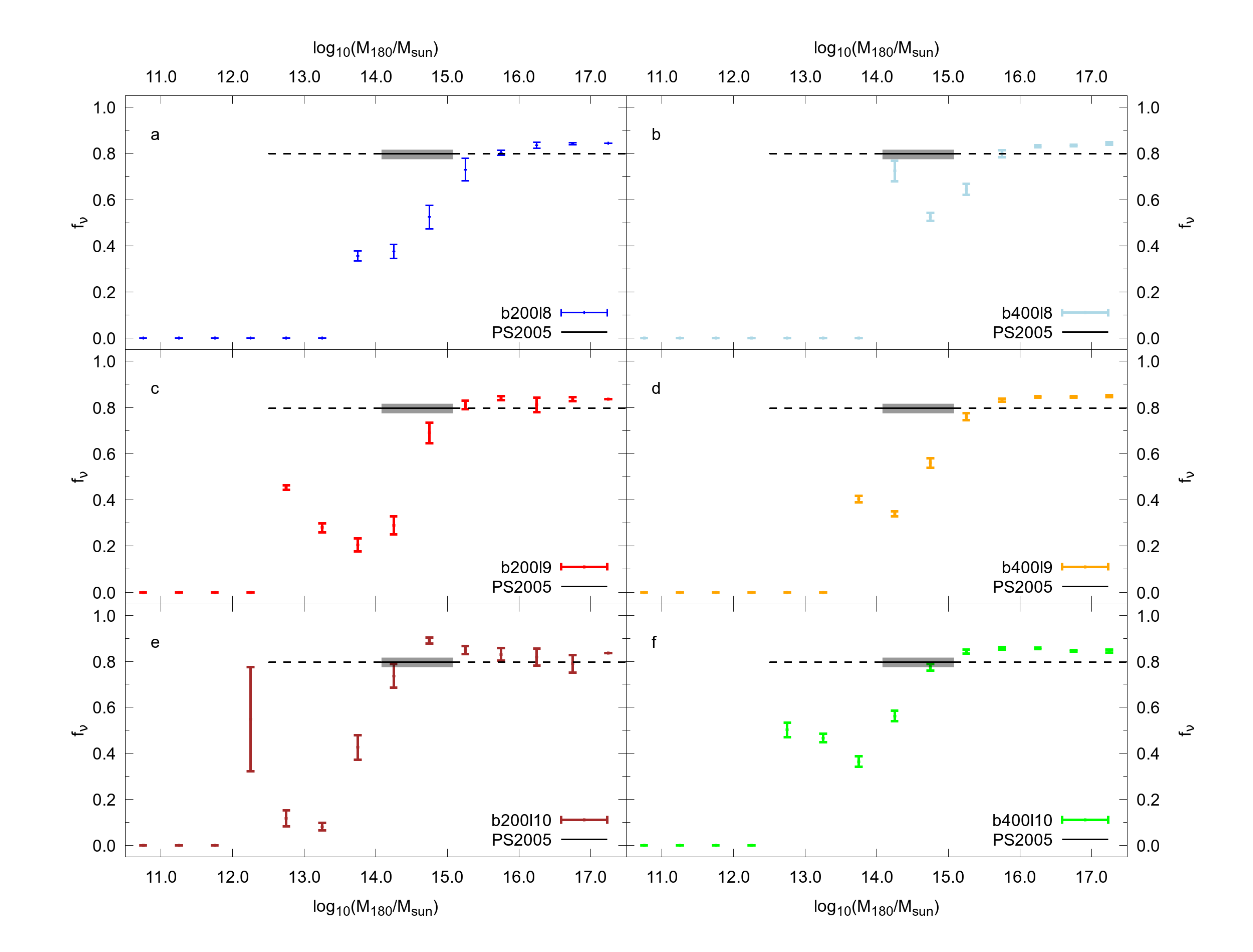}
    \caption{The present ($z_e=0.0$) sterile neutrino mass fraction $f_\nu$ for all models (colours and panel placement are the same as in Figure~\ref{fig:mass-time}). The grey area shows the observed value and its $1\sigma$ uncertainty based on comparing the MOND dynamical mass to the observed baryonic mass of 10 clusters out to $z_e=0.15$ in the mass range $1.2\times 10^{14} - 1.2\times 10^{15} \, M_\odot$ \citep{Pointecouteau_2005}. The dashed line shows the extrapolation of their result outside this mass range. The least massive bound structures identified have masses evident in Figure~\ref{fig:massfunction}.}
    \label{fig:nufraction}
\end{figure*}

These first ever hydrodynamical cosmological MOND simulations with both baryons (gas) and live sterile neutrino particles enable us to explore whether MOND might be consistent with the observed ratio of dynamical mass to baryonic mass, $M_M/M_b$, in massive galaxy clusters, where the ratio appears to be very close to the cosmic fraction \citep[see section~7.1 of][and references therein]{Banik_Zhao_2022}. Since MOND was originally proposed to explain observations with only baryonic mass, we compute the mass fraction of sterile neutrinos, $f_\nu \equiv M_\nu/M_\mathrm{tot}$. Observationally, $M_\mathrm{tot}$ corresponds to $M_M$, the MOND dynamical mass. In our simulations, $M_\mathrm{tot} \equiv M_b + M_\nu$. This can readily be connected to the observed ratio of dynamical mass to baryonic mass: $f_\nu \equiv 1 - M_b/M_M$.

Figure~\ref{fig:nufraction} compares $f_\nu$ for every model (colours and panel placement are the same as in Figure~\ref{fig:mass-time}) with the findings from \citet[][hereafter \citetalias{Pointecouteau_2005}]{Pointecouteau_2005}, who analysed $M_M$/$M_b$ in a MOND context for 10 observed clusters out to a redshift of $z_e=0.15$ in the mass range $1.2\times 10^{14} \,M_\odot$ to $1.2\times 10^{15} \, M_\odot$. They found that $M_M$/$M_b=4.94\pm0.50$ at distances from the cluster center of $r\approx 0.5 \, R_\mathrm{vir}$, in contrast to the SMoC asymptotic ($r \geq 0.3 \, R_\mathrm{vir}$) value of $7.7^{+1.4}_{-1.1}$ (in the SMoC most of the mass is CDM, with neutrinos playing a negligible role). This leads to an observational value for $f_\nu$ of $1 - M_b/M_M = 0.80\pm0.02$, which we show in Figure~\ref{fig:nufraction}. We bin our data in the same mass bins used in Figure~\ref{fig:massfunction} (0.5 dex bin width) and show the mean value with its standard deviation.

It is quite striking that the models come out to be in good agreement with the observational data for $f_\nu$ at the high mass end, especially for the higher resolution models. The slightly higher asymptotic value of all models compared to the observed value is expected because $f_\nu$ for the simulations is calculated at $R_{180}$ but \citetalias{Pointecouteau_2005} evaluated their findings at $r\approx 0.5 \, R_\mathrm{200}$, which is closer to the cluster center. As a result, the simulated value should be closer to the cosmic mean value of $\Omega_\nu/\Omega_m \approx 0.84$, but since the dissipative baryons are expected to be more centrally concentrated than the sterile neutrinos, the neutrino mass fraction should be lower within a smaller radius. The highest resolution model (b200l10; panel e) shows a smooth decline of $f_\nu$ from the cosmic mean value for massive clusters down to a nearly vanishing sterile neutrino component for the least massive structures identified (see Figure~\ref{fig:massfunction}). This nicely shows the transition between clusters and galaxies, where the sterile neutrinos do not cluster by construction in the $\nu$HDM model. Indeed, our results show that the Local Group, with a baryonic mass of $\approx 10^{11} \, M_\odot$, does not have a significant sterile neutrino component in the $\nu$HDM paradigm, in agreement with the past dynamical history of the Milky Way-M31 galaxy pair \citep[the timing argument;][]{Banik_Ryan_2018, Banik_2022_SP, Benisty_2020, McLeod_2020}. More generally, the almost purely baryonic nature of low-mass structures is very much in line with the aim of $\nu$HDM to preserve the empirical successes of MOND in galaxies. Important to note here is that the bins for the lowest masses may be populated by only a few structures, leading to a much larger statistical uncertainty than is shown with the mean value and its standard deviation. This may explain why $f_\nu$ sometimes unexpectedly decreases with increasing mass in the lowest mass bins, but this does not alter the general picture of negligible $f_\nu$ below $10^{14} \, M_\odot$ and $f_\nu \approx 0.84$ above $10^{14} \, M_\odot$. This unexpected behaviour noted here has to be re-evaluated with higher resolution simulations, investigating if this effect depends solely on resolution/mass of the sterile neutrino particles or has an entirely different cause.

\subsection{Feasibility of the observed Local Group peculiar velocity}

\begin{figure*}
    \includegraphics[width=1.0\linewidth]{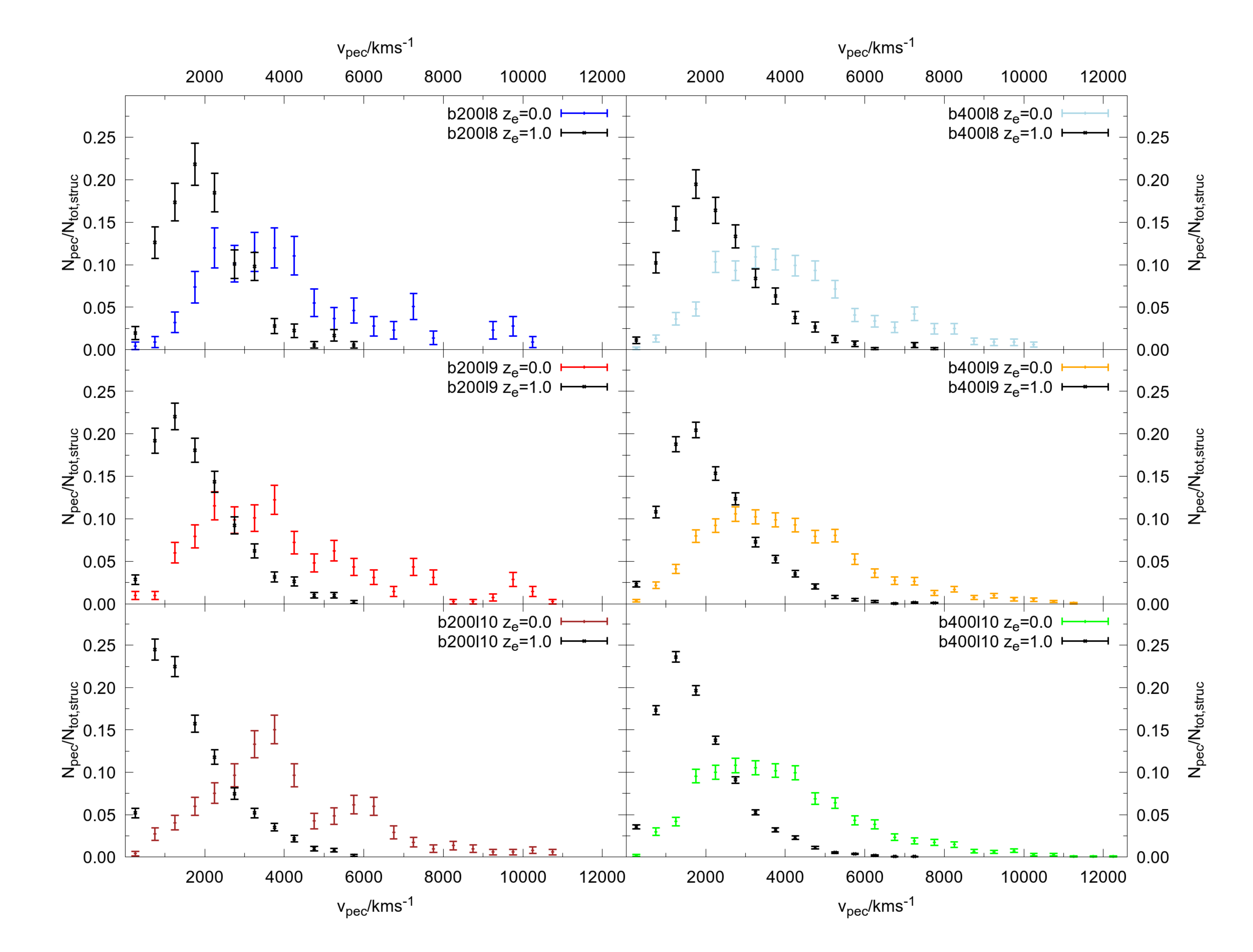}
    \caption{The distribution of the peculiar velocities of all structures at $z_e=1.0$ (black) and at $z_e=0.0$ (coloured according to the model as in Figure~\ref{fig:mass-time}), with a bin width of $500$~km~s$^{-1}$. The number of structures in each velocity bin, $N_\mathrm{pec}$, is normalized by the total number of structures at the relevant time, $N_\mathrm{tot,struc}$. The error bars correspond to Poisson uncertainties.}
    \label{fig:pec_vel}
\end{figure*}

In addition to the mostly mass-related analysis presented here, we show in Figure~\ref{fig:pec_vel} the distribution and time evolution of the peculiar velocities of all structures. We use this to quantify the likelihood of finding a structure with a peculiar velocity as low as that of the Local Group \citep[$v_\mathrm{pec}\leq630$~km~s$^{-1}$;][]{Kogut_1993}.

Figure~\ref{fig:pec_vel} illustrates the peculiar velocity distribution of all structures at $z_e=0.0$ and $z_e=1.0$. The number of structures in each velocity bin of width 500~km~s$^{-1}$ is normalized by the total number of structures, $N_\mathrm{tot,struc}$, at the relevant redshift, with error bars showing the Poisson uncertainty. The black distributions correspond to $z_e=1.0$ and the coloured distributions show the peculiar velocities today. The distributions in different simulations at $z_e=1.0$ are very similar, showing a maximum between 1000~km~s$^{-1}$ and 2000~km~s$^{-1}$ followed by a smooth decline until the highest peculiar velocities are reached at about $v_\mathrm{pec}=6000$~km~s$^{-1}$ for the b200 models and between $7000$ and $8000$~km~s$^{-1}$ for the b400 models. By $z_e=0.0$, the distributions have broadened significantly, shifting the mode and the maximum reached peculiar velocity to larger values. Also, the b200 models show more features in the extended high velocity tail of the distribution, which may be attributed to a smaller number of structures and higher density variations throughout the simulation box (see Figure~\ref{fig:densityz0}). Overall, the peculiar velocities seem to be nearly thermal, i.e., the distributions are comparable to Maxwell-Boltzmann distributions \citep{Maxwell_1877}, as is evident from the asymmetry due to the extended high velocity tail.

Earlier studies of particle-only $\nu$HDM simulations \citep{Katz_2013} show comparable behaviour for the evolution of the median peculiar velocity (see their figure~4). Recent observations of large bulk flow velocities seem to indicate some tension with the SMoC \citep{Watkins_2023}, which may be related to larger than expected peculiar velocities and therefore could be more natural in the $\nu$HDM framework \citep*[see also][]{Vittorio_1986}. Additionally, compatible with the findings from the CMF (see Figure~\ref{fig:cumulative_MF}), the peculiar velocities of the different models exceed those of local structures. In particular, the Local Group has a peculiar velocity of $v_\mathrm{pec}=630$~km~s$^{-1}$ \citep{Kogut_1993}. In the following, we quantify the fraction of structures in each model which have a smaller peculiar velocity. We express this fraction as a tension in terms of the equivalent number of standard deviations (results are presented in Table~\ref{table:pec-vel}).

\begin{table}
    \centering
    \small
    \caption{The likelihood of finding a structure with a peculiar velocity below that of the Local Group ($v_\mathrm{pec}\leq630$~km~s$^{-1}$). The likelihood is shown as a tension in terms of the equivalent number of standard deviations ($\chi$) for a 1D Gaussian, $N_{\mathrm{tot,struc}}$ is the total number of structures in the simulation volume, and $N_{\mathrm{struc,comp}}$ is the number of structures with a sufficiently small peculiar velocity.}
    \begin{tabular}{ccccccc} 
     \hline
     Model & b200 & b400 & b200 & b400 & b200 & b400 \\
           & l10 & l10 & l9 & l9 & l8 & l8 \\
     \hline
     $N_{\mathrm{tot,struc}}$ & 518 & 1509 & 416 & 1467 & 217 & 687\\ 
     $N_{\mathrm{struc,comp}}$ & 3 & 8 & 5 & 10 & 2 & 3\\
     $\chi$ & 2.8 & 2.8 & 2.5 & 2.7 & 2.6 & 2.9\\
     \hline
    \end{tabular}
    \label{table:pec-vel}
\end{table}

\citet{Haslbauer_2020} point out in their section~4.2 that the peculiar velocity of the Local Group is much slower than typically expected in $\nu$HDM such that this observable causes the highest tension between their semi-analytical $\nu$HDM model and the observational constraints. The tension is still only $2.34\sigma$, which is hardly impossible. We find a tension of $2.5\sigma-2.9\sigma$, with the b200 models having a slightly smaller value of $\chi$ than the corresponding b400 models. The models computed for this work show a similar tension to that estimated by \citet{Haslbauer_2020} based on the fraction of their simulated void where the peculiar velocity in the CMB frame is smaller than that observed for the Local Group. Therefore, we conclude that it is unlikely but not impossible to find structures with a comparably low peculiar velocity as the Local Group, which again suggests that the Local Volume is not an average volume of a $\nu$HDM universe $-$ and might also not be for the observed Universe (see Section~\ref{sec:mass-function}).

\section{Conclusions and future work}
\label{Sec:conclusion}

We report the first hydrodynamical MOND simulations of structure formation in a cosmological context. The calculations are performed in the $\nu$HDM cosmological framework proposed by \citet{Angus_2009} in which the CDM component in $\Lambda$CDM is replaced by a collisionless component in the form of sterile neutrinos with a mass of $m_{\mathrm{\nu s}} = 11$~eV/$c^2$. This component is motivated from neutrino physics rather than cosmology but is equivalent to a hot ``dark matter'' component. This conserves the MONDian behaviour on galactic scales but is compatible with galaxy cluster dynamics \citep{Angus_2010} and the CMB \citep{Angus_Diaferio_2011} due to the strong gravitational fields in the early universe and a standard evolution of the cosmic scale factor $a_e$ \citep[for further details, see][]{Haslbauer_2020}. We generate the initial conditions at $z_e=199$ for this cosmological model by calculating the power spectrum with \textsc{camb} while including the necessary changes to only the namelist to replace the CDM component with $11$~eV/$c^2$ sterile neutrinos. We then sample the power spectrum using the \textsc{music} program to generate the initial conditions of the live sterile neutrino particles and the gas distribution. As with other cosmological simulations \citep[e.g.,][]{Candlish_2016}, we solve the Poisson equation in super-comoving coordinates \citep[Equation~\ref{eq:5}; see also Appendix~\ref{appendix:super_co-moving_coords} and][]{Martel_1998, Teyssier_2002}. Applying this to MOND implicitly assumes that the MOND Poisson equation should be applied only to the difference between the density and the cosmic mean value, which we argued to be reasonable in Section~\ref{Sec:simulationcode} \citep[see also section~3.1.4 of][]{Haslbauer_2020}. Recent analytic work in covariant MOND theories further supports this approach \citep{Thomas_2023}.

Six models were computed in this work, exploring the impact of three different resolution settings and two box sizes (Section~\ref{sec:initcond}). We concentrate on small to intermediate sizes ($200 - 400$~cMpc/$h$) in order to resolve the smallest structures possible, while initializing the simulations with enough power to be able to form any structures at all. As the right panel of Figure~\ref{fig:powerspec} depicts, scales below 100~cMpc/$h$ initially have very little power in the matter power spectrum, so starting a simulation with an initial box size in this regime leads to vanishingly small density fluctuations. This would cause structures to arise from numerical noise only, if indeed any form at all. Structure formation must be top-down in the $\nu$HDM framework, so large box sizes are essential.

The first structures begin to form at $z_e \approx 4$ for the higher resolution simulations. This result appears to be in tension with high redshift galaxies observed with the recently launched JWST, which also seem to be in some tension with $\Lambda$CDM \citep{Haslbauer_JWST_2022}. However, previous analytic estimates of structure formation in MOND indicate that galaxy scale structures (up to $10^{11} \, M_\odot$) should have formed by $z_e=5 - 10$, while cluster sized structures ($10^{14} \, M_\odot$) should have reached maximum expansion and begun to recollapse at $z_e=3$ \citep{Sanders_1998_cosmology}. The latter is comparable to the evolution of the models shown here, but important to note is that the analytic estimates are calculated for an idealized spherical scenario without sterile neutrinos and dark energy, representing a rather different framework that may be in tension with the observed CMB. Simulations that adequately resolve galaxy mass scales (rather than just reaching this scale as we do here) are needed to conclusively test the onset of structure formation in the $\nu$HDM framework. We expect a higher resolution simulation to identify the first galaxies at earlier epochs than the structures shown here, with star formation in the first collapsing gas clouds occurring even earlier \citep[see also][]{Wittenburg_2020, Eappen_2022}.

The most massive galaxy clusters (up to $M_{180}=10^{17} \, M_\odot$) form in the models with the largest box size ($b=400$~cMpc/$h$) regardless of resolution (see Figure~\ref{fig:mass-time}). With improved resolution, it may be worthwhile to search for analogues to observed massive galaxies at high redshift \citep[e.g.,][]{Furtak2022, Harikane2022, Labbe2022, Naidu2022a, Adams2023, Artek2023, Yan2023}. If these structures appear in improved simulations with a larger box size, it will become possible to estimate the number density of the most massive galaxy cluster collisions like El Gordo \citep{Asencio_2021} in a statistically meaningful manner.

We also show the differential mass function at $z_e=0$ (Figure~\ref{fig:massfunction}) and compare the simulated CMF with weak lensing data \citep{Wang2022} up to $z_e=0.6$ (Figure~\ref{fig:cumulative_MF}). There is reasonable agreement in the intermediate and low mass regimes, but the slope of the simulated distribution is too shallow, leading to an overabundance of massive structures at the high mass end compared to the weak lensing catalogue. The overabundance is similar to that mentioned in \citet{Angus_2013} and \citet{Katz_2013}, but we argue that, especially for the highest masses, the relatively local observations might not be representative of an average volume of the Universe. In particular, we are situated in a supervoid with a radius of $r\approx300$~Mpc \citep[the KBC void;][]{Keenan_2013, Haslbauer_2020}, which inhibits the existence of very massive clusters as the most massive structures can only form in the most dense regions of the universe. Additionally, observations out to $z_e\approx2$ \citep{Dainotti_2021, Dainotti_2022, Jia2023} show a smooth decrease of the inferred $H_0$ from the high local value \citep{Riess2019} towards the smaller value inferred from the CMB \citep{Planck_2020}. Assuming that the tension between the two values of $H_0$ vanishes when considering a large underdensity in MOND \citep{Haslbauer_2020}, one interpretation of a smooth transition between the two values might be that the underdensity around us is larger than shown by the KBC void (the inference on the void radius in their figure~4 does indeed show an extended tail towards large values). This would lead to the universe not being as homogeneous as predicted by the SMoC. There are indeed observed structures like El Gordo which are at relatively high redshift ($z_e=0.87$) and are extremely massive \citep[$M_{200}=2\times10^{15} \, M_\odot$;][]{Kim_2021}, with El Gordo alone ruling out the $\Lambda$CDM cosmology at $>6\sigma$ confidence \citep{Asencio_2021}. We therefore argue that the $\nu$HDM model cannot yet be ruled out by the mismatch between observations and simulations at the high mass end of the cluster mass function.

To improve our understanding of this issue, a more detailed analysis would be required to check if the galaxy or galaxy cluster mass function for an observer in a low-density region matches the observations and if large volume (Gpc scale) simulations can reproduce El Gordo analogues with the observed frequency. As a first step towards this, we divide the b200l10 model into 8 equally sized octants and compare their mass functions at $z_e=0$ and $z_e=1$. Figure~\ref{fig:massfunction100Mpc} shows that there is more than one dex difference in density between the sub-volumes and also that the most massive clusters only form in the densest regions. The figure also shows that our models form clusters of the same mass scale as the progenitors of El Gordo at $z_e=1$, indicating that $\nu$HDM could be consistent with the observations of El Gordo \citep[as argued in section~4.3 of][]{Asencio_2021}.

We also examine the time evolution of the structures by showing how the number of structures and the baryonic mass inside all structures evolve with time (Figures~\ref{fig:N_time} and \ref{fig:mfraction}, respectively). The former yields the first estimates for the prevalence of mergers in a cosmological MOND framework, while also showing that this seems to be resolution dependent. The latter, in combination with the former, suggests that the accretion of gas on to structures dominates the growth of mass in structures for $z_e<2$ in these models.

Additionally, we are for the first time able to show the mass ratio of baryons to sterile neutrinos in bound structures in the form of the sterile neutrino mass fraction $f_\nu$, which we compare to the observed ratio of MOND dynamical mass to baryonic mass \citepalias{Pointecouteau_2005}. Figure~\ref{fig:nufraction} shows that the higher resolution simulations are consistent with the observations in the observed mass range. In particular, $f_\nu$ becomes asymptotically flat at the cosmic value of $\Omega_\nu/\Omega_m\approx0.84$ for structures more massive than $10^{15} \, M_\odot$, a result that is only weakly dependent on resolution and box size. The highest resolution simulation (panel e) confirms the expected $\nu$HDM behaviour that the fraction of mass in sterile neutrinos decreases from cluster mass scales down to galaxy mass scales, where it vanishes. This is required in any MOND cosmology because galaxies can be explained purely by their baryonic mass content in MOND \citep[][and references therein]{Famaey_McGaugh_2012, McGaugh_2020, Brouwer_2021, Banik_Zhao_2022}.

Furthermore, we show the distribution of peculiar velocities of all structures and calculate the likelihood of finding structures with a peculiar velocity as small as that of the Local Group. We find this to cause between $2.5\sigma$ and $2.9\sigma$ tension depending on the model, with the b200 models having a slightly smaller tension compared to their b400 counterparts (see Table~\ref{table:pec-vel}). Our result is compatible with the findings of \citet{Haslbauer_2020}, according to which a $2.34\sigma$ tension exists between observations and their semi-analytical $\nu$HDM model in this respect. This shows that a peculiar velocity as low as the observed 630~km/s is unlikely but not impossible in $\nu$HDM. Therefore, the Local Volume is rare in a $\nu$HDM universe and perhaps also in the real Universe, as discussed further in Section~\ref{sec:mass-function}. We also note that the large observed bulk flow out to $\approx 200$~Mpc is in some tension with $\Lambda$CDM \citep{Watkins_2023} and may be a sign of the typically much larger peculiar velocities in $\nu$HDM \citep[see our Figure~\ref{fig:pec_vel} and also figure~4 of][]{Katz_2013}.

In this work, we present the first hydrodynamical cosmological MOND simulations. This is an important step towards a more detailed investigation of structure formation in a MOND cosmology. Further simulations similar to \citet{Katz_2013} are needed to analyse the behaviour of this model on larger scales, where it appears to provide a promising explanation for the KBC void and Hubble tension \citep{Haslbauer_2020} and the massive high-redshift interacting galaxy cluster known as El Gordo \citep{Asencio_2021}, phenomena which contradict $\Lambda$CDM at high significance. Moreover, the scales considered here need to be simulated at higher resolution with star formation and feedback prescriptions to investigate how single galaxies form and evolve in this model, which will be the task for future works. We are also conducting larger volume collisionless simulations to better explore the behaviour of our model on the largest scales (Russell et al., in preparation). We want to stress that improved simulations will be essential to check if the shallow mass function and the late onset of structure formation prevail in higher resolution simulations in the $\nu$HDM framework, which would falsify this model conclusively.

\section*{Acknowledgements}

IB is supported by Science and Technology Facilities Council grant ST/V000861/1. He acknowledges support from a ``Pathways to Research'' fellowship from the University of Bonn in 2021 after an Alexander von Humboldt Foundation postdoctoral research fellowship ($2018-2020$). We thank the Deutscher Akademischer Austauschdienst-Eastern European exchange program at the University of Bonn for supporting the Bonn-Prague exchange and acknowledge financial support through the transdisciplinary research area TRA-Matter at the University of Bonn. The authors thank Benoit Famaey and Alfie Russell for useful discussions. They also thank the anonymous referee for helpful comments.

\section*{Data availability}

The algorithms used to prepare and run \textsc{por} simulations in a cosmological context and to extract their results into human-readable form are publicly available.\footnote{\url{https://bitbucket.org/SrikanthTN/bonnpor/src/master/Cosmo_patch_and_setup_and_halofinder/}} \citet{Nagesh_2021} provides a user guide describing the operation of these algorithms.

\bibliographystyle{mnras}
\bibliography{vHDM_setup_bbl}

\begin{appendix}

\section{The Poisson equation in super-comoving coordinates}
\label{appendix:super_co-moving_coords}

Following the derivation of \citet{Martel_1998}, we start with their eq. 14, without the addition of the non-clumping X-component that they use:
\begin{eqnarray}
\label{eq:Poisson_beginning}
    \nabla^2\Phi &=& 4\mathrm{\pi} G \rho \, ,
\end{eqnarray}
where $\Phi$ is the Eulerian gravitational potential \citep[eq.~23 in][]{Martel_1998},
\begin{eqnarray}
\label{eq:Phi_Eulerian}
    \Phi &=& \frac{2\mathrm{\pi} G \bar{\rho} r^2}{3} + \phi \, ,
\end{eqnarray}
with $\bar{\rho}$ being the mean matter density, i.e., the left part of the right hand side represents the background potential and $\phi$ the peculiar gravitational potential. We want $\phi$ as we need to apply the \textit{Jeans Swindle} \citep{Binney_Tremaine_1987, Binney_Tremaine_2008}, so we differentiate Eq.~\ref{eq:Phi_Eulerian} two times and additionally use Eq.~\ref{eq:Poisson_beginning} to get the Poisson equation for $\phi$:
\begin{eqnarray}
    \nabla^2 \Phi &=& 4\mathrm{\pi} G \bar{\rho} + \nabla^2\phi = 4\mathrm{\pi} G \rho \, \\
    \label{eq:Poisson_cosmo}
    \Leftrightarrow \nabla^2 \phi &=& 4 \mathrm{\pi} G \left(\rho-\bar{\rho}\right) \, .
\end{eqnarray}
This is the Poisson equation for cosmological calculations, which only applies to density contrasts rather than the total density field.

Now, we want to transform Eq.~\ref{eq:Poisson_cosmo} into super-comoving coordinates as these conserve the positions, the density, and the thermodynamic variables of an ideal gas in a cosmologically expanding volume in the absence of structure. To achieve this, several variables need to be transformed. We use the transformation from \citet{Martel_1998}:\footnote{Further variables need to be transformed in order to completely transform the Euler equations plus the Poisson equation, but the Euler equations have the same form after the transformation \citep[e.g.,][]{Teyssier_2002}.}:
\begin{eqnarray}
    \label{eq:r_tilde}
    \Tilde{r} &=& \frac{r}{ar_\star} \, ,\\
    \label{eq:rho_tilde}
    \Tilde{\rho} &=& \frac{a^3\rho}{\rho_\star} \, ,\\
    \label{eq:rho_star}
    \rho_\star &=& \frac{3 H_0^2 \Omega_{m,0}}{8 \mathrm{\pi} G} \, ,\\
    \label{eq:t_tilde}
    d\Tilde{t} &=& \frac{dt}{a^2t_\star} \, ,\\
    \label{eq:t_star}
    t_\star &=& \frac{1}{H_0} \, ,\\
    \label{eq:phi_tilde}
    \Tilde{\phi} &=& \frac{a^2\phi}{\phi_\star} \, ,\\
    \label{eq:phi_star}
    \phi_\star &=& \frac{r_\star^2}{t_\star^2} \, ,
\end{eqnarray}
where $\Omega_{m,0}\equiv\bar{\rho}/\rho_{crit}$, the critical density $\rho_{crit}\equiv3H_0^2/8\mathrm{\pi} G$, $\rho$ is the density, $t$ is the time, $\phi$ is the gravitational potential, and $\bm{r}$ is the spatial position. The tilde symbol denotes supercomoving coordinates. Every parameter with a star subscript is a necessary fixed parameter for the transformation, with $r_\star$ being the only independent parameter $-$ this is usually taken to be the simulation box size. Note that we use the $t_\star$ definition from \citet{Teyssier_2002}. We also need to convert the derivatives relative to $r$ at fixed $t$ to derivatives relative to $\Tilde{r}$ at fixed $\Tilde{t}$,
\begin{eqnarray}
\label{eq:derivative}
    \left(\nabla f\right)_t &=& \frac{1}{a r_\star}\left(\Tilde{\nabla} f\right)_{\Tilde{t}} \, ,
\end{eqnarray}
where $\Tilde{\nabla}$ is the gradient relative to $\Tilde{r}$ and $f$ is an arbitrary function, which was not transformed here. The reader should keep in mind that if $f$ depends on parameters that are transformed, then $f$ also needs to be transformed accordingly.

We can start inserting the transformation now:
\begin{eqnarray}
    \nabla^2 \phi &=& 4 \mathrm{\pi} G \left(\rho-\bar{\rho}\right) \, ,\\
    \stackrel{\text{\ref{eq:phi_tilde}}}{\Leftrightarrow} \nabla^2\left(\frac{\Tilde{\phi} \phi_\star}{a^2}\right) &=& 4 \mathrm{\pi} G \left(\rho-\bar{\rho}\right) \, ,\\
    \stackrel{\text{\ref{eq:derivative}}}{\Leftrightarrow} \frac{\phi_\star}{a^4 r_\star^2} \Tilde{\nabla}^2\left(\Tilde{\phi}\right) &=& 4 \mathrm{\pi} G \left(\rho-\bar{\rho}\right) \, ,\\
    \stackrel{\text{\ref{eq:phi_star}}}{\Leftrightarrow} \frac{1}{a^4 t_\star^2} \Tilde{\nabla}^2\left(\Tilde{\phi}\right) &=& 4 \mathrm{\pi} G \left(\rho-\bar{\rho}\right) \, ,\\
    \stackrel{\text{\ref{eq:rho_tilde}}}{\Leftrightarrow} \frac{1}{a^4 t_\star^2} \Tilde{\nabla}^2\left(\Tilde{\phi}\right) &=& \frac{4 \mathrm{\pi} G}{a^3} \left(\Tilde{\rho}\rho_\star - \stackrel{\tiny\text{\texttildelow}}{\bar{\rho}}\rho_\star\right) \, ,\\
    \stackrel{\text{\ref{eq:rho_star}}}{\Leftrightarrow} \frac{1}{a^4 t_\star^2} \Tilde{\nabla}^2\left(\Tilde{\phi}\right) &=& \frac{3 H_0^2 \Omega_{m,0}}{2 a^3} \left(\Tilde{\rho} - \stackrel{\tiny\text{\texttildelow}}{\bar{\rho}}\right) \, ,\\
    \Leftrightarrow \Tilde{\nabla}^2\left(\Tilde{\phi}\right) &=& 3 a H_0^2 t_\star^2 \Omega_{m,0} \left(\Tilde{\rho} - \stackrel{\tiny\text{\texttildelow}}{\bar{\rho}}\right) \, ,\\
    \stackrel{\text{\ref{eq:t_star}}}{\Leftrightarrow} \Tilde{\nabla}^2\left(\Tilde{\phi}\right) &=& 3a \Omega_{m,0}\left(\Tilde{\rho} - \stackrel{\tiny\text{\texttildelow}}{\bar{\rho}}\right) \, .
\end{eqnarray}
For QUMOND, $\rho_{tot}$ is effectively $\rho+\rho_p$, with $\rho_p$ in physical units:
\begin{eqnarray}
    \rho_p ~=~ \frac{1}{4 \mathrm{\pi} G} \nabla \cdot \left[\Tilde{\nu}\left(\frac{\lvert\nabla\phi_N\rvert}{a_0}\right)\nabla\phi_N\right].
\end{eqnarray}
Since $\Tilde{\rho}_p=a^3\rho_p/\rho_\star$, we get $\rho_p$ in super-comoving units using again Eqs.~\ref{eq:rho_star}, \ref{eq:t_star}, \ref{eq:phi_tilde}, \ref{eq:phi_star}, and \ref{eq:derivative}:
\begin{eqnarray}
    \Tilde{\rho}_p &=& \frac{2}{3 a \Omega_{m,0}} \Tilde{\nabla} \cdot \left[\Tilde{\nu}\left(\frac{\lvert\nabla\phi_N\rvert}{a_0}\right)\Tilde{\nabla}\Tilde{\phi}_N\right].
\end{eqnarray}
Important to note is that the argument of $\Tilde{\nu}$, i.e., $\lvert\nabla\phi_N\rvert/a_0$, is dimensionless and therefore not affected by this coordinate transformation.

Finally, the Poisson equation for QUMOND in super-comoving coordinates is:
\begin{eqnarray}
    \Tilde{\nabla}^2\left(\Tilde{\phi}\right) &=& 3a\Omega_{m,0} \left(\Tilde{\rho} - \stackrel{\tiny\text{\texttildelow}}{\bar{\rho}}\right)\\
    &+& \Tilde{\nabla} \cdot \left[\Tilde{\nu}\left(\frac{\lvert\nabla\phi_N\rvert}{a_0}\right)\Tilde{\nabla}\Tilde{\phi}_N\right] \nonumber.
\end{eqnarray}

\end{appendix}

\bsp
\label{lastpage}
\end{document}